\documentclass[11pt, letterpaper]{article}

\usepackage[USenglish]{babel}
\usepackage{tikz}
\usepackage{graphicx}
\usepackage[ansinew]{inputenc}
\usepackage[T1]{fontenc}
\usepackage{amsmath}
\usepackage{enumitem}
\usepackage{amssymb}
\usepackage{amsthm}
\usepackage{mathtools}
\usepackage{booktabs}
\usepackage{hyperref}
\usepackage[left=2.5cm,right=2.5cm,top=2cm,bottom=2cm,includeheadfoot]{geometry}
\usepackage{xspace}
\usepackage{thmtools}
\usepackage{thm-restate}
\usepackage{float}
\usetikzlibrary{arrows.meta,arrows}
\usetikzlibrary{positioning,arrows}
\usetikzlibrary{decorations.pathreplacing}
\usepackage{cleveref}
\usepackage{footmisc}
\usepackage{stackrel}
\allowdisplaybreaks

\usepackage{todonotes}
\usepackage{authblk}

\def\myoverset#1#2#3{\stackrel{\text{\makebox[#3pt]{#1}}}{#2}}
\def\mybothset#1#2#3#4#5{\stackrel[\text{\makebox[#5pt]{#3}}]{\text{\makebox[#4pt]{#1}}}{#2}}

\theoremstyle{definition}
\newtheorem{Sat}{Satz}
\numberwithin{Sat}{section}

\newtheorem{Kor}[Sat]{Corollary}

\newtheorem{Def}[Sat]{Definition}
\newtheorem{Obs}[Sat]{Observation}
\newtheorem*{NBG}{Normal Block Gadget}
\newtheorem*{OBG}{Origin Block Gadget}
\newtheorem*{FBG}{Closing Block Gadget}
\newtheorem*{NBGO}{Normal Block Gadget of Level $0$}
\newtheorem*{OBGO}{Origin Block Gadget of Level $0$}
\newtheorem*{FBGO}{Final Block Gadget of Level $0$}
\newtheorem*{NBGI}{Normal Block Gadget of Level $0<i<N$}
\newtheorem*{OBGI}{Origin Block Gadget of Level $0<i<N$}
\newtheorem*{FBGI}{Final Block Gadget of Level $0<i<N$}
\newtheorem*{NBGN}{Normal Block Gadget of Level $N$}
\newtheorem*{OBGN}{Origin Block Gadget of Level $N$}
\newtheorem*{FBGN}{Final Block Gadget of Level $N$}
\newtheorem*{CBGN}{Closing Block Gadget of Level $N$}

\newtheorem*{GTwo}{Graph $G_\text{simple}$}
\newtheorem*{GThree}{Graph $G_\text{rec}$}
\newtheorem*{GFour}{Graph $G_\text{chain}$}

\newcommand{\uX}[1]{u_{#1}^\text{rec}}
\newcommand{\ucX}[1]{u_{#1}^\text{chain}}
\newcommand{\utX}[1]{\tilde{u}_{#1}}
\newcommand{\vX}[1]{o_{#1}}
\newcommand{\eX}[1]{e_{#1}}
\newcommand{\tX}[1]{t_{#1}}

\newcommand{\opt}{\textsc{Opt}\xspace}
\newcommand{\alg}{\textsc{Alg}\xspace}

\newcommand{\N}{\mathbb{N}}

\newboolean{comments}
\setboolean{comments}{false}
\ifthenelse{\boolean{comments}}{
	\newcommand{\alexB}[1]{\textcolor{blue}{\boldall{Alex B.: #1}}}
	\newcommand{\alexH}[1]{\textcolor{green!65!black}{\boldall{Alex H.: #1}}}
	\newcommand{\yannD}[1]{\textcolor{orange}{\boldall{Yann: #1}}}
	\newcommand{\christinaK}[1]{\textcolor{red}{\boldall{Christina: #1}}}
}{
	\newcommand{\alexB}[1]{}
	\newcommand{\alexH}[1]{}
	\newcommand{\yannD}[1]{}
	\newcommand{\christinaK}[1]{}
}

\newboolean{full}
\setboolean{full}{true}
\ifthenelse{\boolean{full}}{
	\newcommand{\full}[1]{#1}
	\newcommand{\conf}[1]{}
}
{
	\newcommand{\full}[1]{}
	\newcommand{\conf}[1]{#1}
}

\newboolean{shorten}
\setboolean{shorten}{true}
\ifthenelse{\boolean{shorten}}{
	\newcommand{\shorten}[1]{}
}
{
	\newcommand{\shorten}[1]{#1}
}

\newcommand{\captionshift}{\vspace*{-1em}}

\author[1]{Alexander Birx}
\author[1]{Yann Disser}
\author[1]{Alexander V. Hopp}
\author[1]{Christina Karousatou}
\affil[1]{
	\small Department of Mathematics, TU Darmstadt, Germany.
}

\begin{document}

\title{Improved Lower Bound for\\ Competitive Graph Exploration\full{\thanks{This work was supported by the `Excellence Initiative' of the German Federal and State Governments and the Graduate School~CE at TU~Darmstadt.}}}
\date{}
\maketitle

\begin{abstract}
  We give an improved lower bound of~$\frac{10}{3}$ on the competitive ratio for the exploration of an undirected, edge-weighted graph with a single agent that needs to return to the starting location after visiting all vertices.
  We assume that the agent has full knowledge of all edges incident to visited vertices, and, in particular, vertices have unique identifiers.
  Our bound improves a lower bound of~$2.5$ by Dobrev et al.~[SIROCCO'12] and also holds for planar graphs, where it complements an upper bound of~$16$ by~Kalyanasundaram and Pruhs~[TCS'94]\yannD{Nicole?}.
  The question whether a constant competitive ratio can be achieved in general remains open.
\end{abstract}

\section{Introduction}

We consider the problem of exploring an initially unknown, undirected and edge-weighted graph with an agent starting at some vertex called the \emph{origin}.
In each step, the agent may move along any edge incident to its current location, which incurs a cost equal to the weight of the edge.
The agent has full knowledge of all edges incident to visited vertices, including their weights and the identities of any unvisited vertices they may lead to.
In this setting, the objective of the agent is to visit all vertices and return to the origin while minimizing the total cost of edge traversals.
We measure the performance of an exploration algorithm by its competitive ratio, i.e., the worst case ratio between its cost and the cost of an optimum offline solution that knows the graph beforehand. \shorten{, which is equal to the cost of an optimum TSP-tour.}

This problem was first introduced by Ka\-ly\-a\-na\-sun\-da\-ram and Pruhs~\cite{Kalyanasundaram1994}, who gave a 16-com\-pe\-ti\-tive algorithm for planar graphs.\footnote[1]{Note that early works referred to the exploration problem as ``online TSP'', a name now reserved for the problem where the graph is known but the vertices that need to be visited are revealed over time~(see~\cite{BjeldeDisserHackfeldEtal/17}).}
Megow et al.~\cite{MegowMehlhornSchweitzer/11} showed that this algorithm has a bounded competitive ratio only on graphs of bounded genus.
The best known lower bound for the weighted case is $2.5$ and was given by Dobrev et al.~\cite{Dobrev2012}; it holds even for planar graphs.
The question whether a constant competitive ratio can be achieved in general remains open.

\paragraph*{Our Results.}

We improve the lower bound of Dobrev et al.~\cite{Dobrev2012} from $2.5$ to $\frac{10}{3}$.
To do this, we first tweak their construction within each recursive layer to obtain an improved bound of~$3$ and then modify the resulting graph macroscopically to further improve this bound to~$\frac{10}{3}$.
Note that our bound still holds for planar graphs, where the best known upper bound is~$16$, due to Kalyanasundaram and Pruhs~\cite{Kalyanasundaram1994}.
We hope that our construction gives some helpful insights towards the question whether a constant competitive ratio is attainable.

\paragraph*{Related Work.}

A natural exploration algorithm is the nearest-neighbor algorithm that always explores the unexplored vertex that is cheapest to reach for the agent.
This algorithm is a well-known heuristic for TSP and is $\Theta(\log n)$-competitive~\cite{Rosenkrantz1977}.
Hurkens and Woeginger~\cite{Hurkens2004} showed that the lower bound for the competitive ratio of this heuristic already holds for unweighted, planar graphs.
Note that, for unweighted graphs, DFS is obviously 2-competitive, and Miyazaki et al.~\cite{Miyazaki2009} showed that this is best-possible.
Improved bounds are known for special classes of (weighted) graphs~\cite{Asahiro2010, Brandt2019, Miyazaki2009}. 
Most prominently, Miyazaki et al.~\cite{Miyazaki2009} gave a $(1+\sqrt{3})/2$-competitive algorithm for cycles and showed that this is best-possible.

For exploring all vertices of a directed graph, Foerster and Wattenhofer~\cite{Foerster2016} gave upper and lower bounds on the best-possible competitive ratio that are linear in the number of vertices.
The problem changes significantly if we require all edges to be explored~\cite{Albers2000, Deng1999, Fleischer2005}. 
Deng and Papadimitriou~\cite{Deng1999} showed that the best competitive ratio for this setting depends on the deficiency~$d$ of the graph, i.e., the number of edges we need to add to make the graph Eulerian.
The best known algorithm is due to Fleischer and Trippen~\cite{Fleischer2005} and has competitive ratio $\mathcal{O}(d^8)$.

For collaborative exploration with teams of agents, several bounds have been shown for undirected, unweighted graphs~\cite{Dereniowski2013, Disser/18, Dynia2007, Fraigniaud2006, Higashikawa2012, Ortolf2014}.
A constant competitive ratio can be achieved for teams of constant or exponential size by using DFS or BFS, respectively.
Dereniowski et al.~\cite{Dereniowski2013} showed that a constant competitive ratio is possible already (roughly) for a quadratic number of agents.
Tight results are not yet known for smaller, non-constant teams (see~\cite{Disser/18} for a survey).

Exploration of undirected, unweighted graphs has also been studied from an information perspective, usually with the assumption that vertices do not have unique identifiers.
For example, the tradeoff between the size of \emph{advice} available to the algorithm and its performance has been studied~\cite{Bockenhauer2018, Dobrev2012, Gorain2018}.
Also, it is known that~$\Theta(\log n)$ memory bits are needed for exploration~\cite{Fraigniaud2005, Reingold2008}, and this number can be lowered if the agent has~$\Theta(\log \log n)$ pebbles available to it~\cite{Disser2016}.

\paragraph*{Outline.}

We first describe the basic construction with a single layer in Section~\ref{section: Basic Construction}.
This construction proceeds along the same lines as the construction by Dobrev et al.~\cite{Dobrev2012} but introduces an adaptation that allows us to obtain an improved factor of~$3$ when applying this construction recursively in Section~\ref{section: Improved Recursive Construction}.
In Section~\ref{section: Further Improvements to the Lower Bound} we introduce an additional modification of the macroscopic structure\shorten{ of our construction} that\shorten{ further} improves our result to a lower bound of~$\frac{10}{3}$\shorten{ on the competitive ratio}.
Finally, in Section~\ref{section: Conclusion}, we discuss some properties of our construction regarding planarity and the number of different weights involved.

\section{Lower Bound of \texorpdfstring{$2$}{2}}\label{section: Basic Construction}

Let $\alg$ be a fixed deterministic algorithm for solving the graph exploration problem. 
We write $\alg(G)$ for the costs \alg accumulates when traversing an edge-weighted graph $G$.
Similarly, we write $\opt(G)$ for the costs of the optimum traversal of $G$.
The basic idea is to construct a graph $G_{\text{simple}}$ with origin $v_o$ consisting of a cycle of \emph{block gadgets} (see \Cref{figure: Simple Circle Overview}).
These blocks are constructed in a way that \opt can traverse them cheaply, while \alg traverses them inefficiently the first time it visits them. 
\alg adds up roughly three times the cost of \opt for traversing a block the first time.
However, both incur the same cost for any traversal of an explored block and for the transition between the blocks.
We will prove the following result in this section.

\newcommand{\FigureSimpleCircleOverview}{
\begin{figure}\conf{[p]}
\begin{center}
\begin{tikzpicture}[scale=0.25,font=\small]
\draw[line width=2pt,black,-] (-5,0)--(-22.5,0)--(-22.5,10)--(22.5,10)--(22.5,0)--(5,0);
\draw (0,0) node {\huge ...};

\draw[line width=1.5pt,red,->, >=stealth] (0.5,8)--(19.5,8)--(19.5,2)--(-19.5,2)--(-19.5,8)--(-0.5,8);
\draw[line width=1.5pt,blue,->, >=stealth] (0.5,7)--(18.5,7)--(18.5,3)--(-18.5,3)--(-18.5,7)--(-0.5,7);

\draw[fill=white, line width=1pt] (-24.5,11) rectangle (-20.5,9);
\draw[fill=white, line width=1pt] (-24.5,-1) rectangle (-20.5,1);
\draw[fill=white, line width=1pt] (24.5,11) rectangle (20.5,9);
\draw[fill=white, line width=1pt] (24.5,-1) rectangle (20.5,1);

\draw[fill=white, line width=1pt] (-17,11) rectangle (-13,9);
\draw[fill=white, line width=1pt] (-17,-1) rectangle (-13,1);
\draw[fill=white, line width=1pt] (17,11) rectangle (13,9);
\draw[fill=white, line width=1pt] (17,-1) rectangle (13,1);

\draw[fill=white, line width=1pt] (-9.5,11) rectangle (-5.5,9);
\draw[fill=white, line width=1pt] (-9.5,-1) rectangle (-5.5,1);
\draw[fill=white, line width=1pt] (9.5,11) rectangle (5.5,9);
\draw[fill=white, line width=1pt] (9.5,-1) rectangle (5.5,1);

\draw[fill=white, line width=1pt] (-24.5,4) rectangle (-20.5,6);
\draw[fill=white, line width=1pt] (24.5,4) rectangle (20.5,6);

\draw[fill=white, line width=1pt] (-2,11) rectangle (2,9);
\end{tikzpicture}
\end{center}
\captionshift
\caption{The basic design of the graph of instance $G$. The blocks can only be explored inefficiently by \alg. In blue the route of \opt, in red a possible route of \alg.} \label{figure: Simple Circle Overview}
\end{figure}
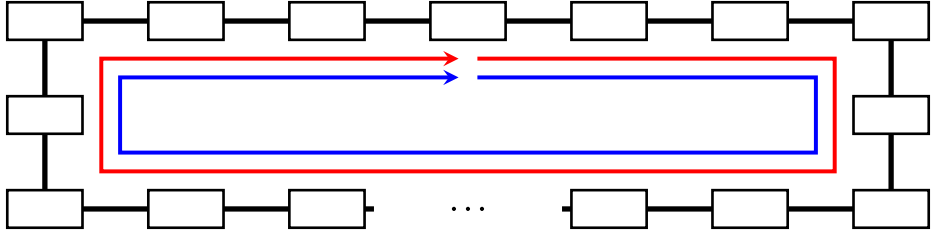
}
\full{\FigureSimpleCircleOverview}

\begin{restatable}{The}{TheoremLowerBoundTwo}\label{theorem: Lower Bound 2}
For every $\varepsilon>0$, there is a graph $G_\text{simple}$ such that $\alg(G_\text{simple})\ge (2-\varepsilon)\cdot\opt(G_\text{simple})$.
\end{restatable}

\subsection{Basic Block Strategy}\label{subsection: Basic Block Strategy}

We start by describing block gadgets.
In the following, $x\in\N$ is a sufficiently large number.
We construct the blocks in a way that exploring one block and entering another block afterwards can be done for an optimum cost of only $2x$, while \alg incurs a cost of at least $4x-1$. 

See \Cref{figure: Normal Block} for an illustration of a normal block gadget (defined below). Note that this gadget is similar to the one in~\cite{Dobrev2012}. 
However, importantly, in our gadget, the position of the start vertex~$v_\text{start}$, which is the first vertex inside of the gadget that is visited by \alg, is variable and not fixed. 

\newcommand{\figureNormalBlock}{
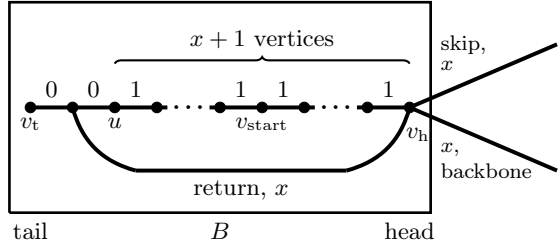
\begin{figure}\conf{[p]}
\begin{center}
\begin{tikzpicture}[scale=0.28,font=\footnotesize,decoration=brace]
\draw[line width=1pt,black,-] (-30,0)--(-10,0)--(-10,10)--(-30,10)--(-30,0);

\draw (-29,5) node[below] {$v_{\text{t}}$};
\draw (-25,5) node[below] {$u$};
\draw (-28,5) node[above] {$0$};
\draw (-26,5) node[above] {$0$};
\draw (-17,5) node[above] {$1$};
\draw (-19,5) node[above] {$1$};
\draw (-18,5) node[below] {$v_{\text{start}}$};
\draw (-24,5) node[above] {$1$};
\draw (-12,5) node[above] {$1$};
\draw (-10.6,4.5) node[below] {$v_{\text{h}}$};

\draw (-10,8) node[right] {\scriptsize skip,};
\draw (-10,7) node[right] {\scriptsize $x$};
\draw (-10,3) node[right] {\scriptsize $x,$};
\draw (-10,2) node[right] {\scriptsize backbone};

\draw (-19,2) node[below] {return, $x$};

\draw (-20,0) node[below] {$B$};

\draw (-29,0) node[below] {tail};
\draw (-11,0) node[below] {head};

\foreach \x in{-29,-27,-25,-23,-20,-18,-16,-13,-11}{
	\filldraw[black] (\x,5) circle (0.25);
}

\filldraw[white] (-40,5) circle (0.25);
\filldraw[white] (0,5) circle (0.25);

\draw (-21.5,5) node {\LARGE ...};
\draw (-14.5,5) node {\LARGE ...};

\draw[line width=1.5pt, black, -] (-29,5)--(-22.5,5);
\draw[line width=1.5pt, black, -] (-20.5,5)--(-15.5,5);
\draw[line width=1.5pt, black, -] (-13.5,5)--(-11,5);

\draw[line width=1.5pt, black, -] (-11,5) -- (-4,8);
\draw[line width=1.5pt, black, -] (-11,5) -- (-4,2);

\draw[line width=1.5pt, black, -] (-11,5) to[bend left] (-14,2);
\draw[line width=1.5pt, black, -] (-24,2)--(-14,2);
\draw[line width=1.5pt, black, -] (-27,5) to[bend right] (-24,2);

\draw[decorate, yshift=2ex, thick]  (-25,6.75) -- node[above=0.4ex] {$x+1$ vertices}  (-11,6.75);
\end{tikzpicture}
\end{center}
\captionshift
\caption{A normal block. 
The position of  $v_\text{start}$ depends on the actions of \alg.} \label{figure: Normal Block}
\end{figure}
}
\full{\figureNormalBlock}

\begin{NBG}\label{strategy: Simple Block}
Assume \alg starts at some vertex, which we call start vertex $v_{\text{start}}$, and has two incident edges of weight~$1$ to choose from.
Every time \alg chooses to traverse an unexplored edge of weight~$1$, it arrives at a new vertex that is incident to another edge of weight~$1$.
This way, \alg explores the graph and every new vertex it visits is incident to exactly one unexplored edge of weight~$1$. 
We stop this procedure once \alg has explored~$x$ vertices and call the current position of \alg \emph{head vertex} $v_\text{h}$.
The head vertex is incident to three edges of weight~$x$.
\alg now has the choice between backtracking to the first unvisited vertex $u$ to the other side of the path or taking one of the three edges of weight $x$. 
If \alg chooses to traverse an edge of weight $x$, we call it a \textit{return edge} (the other two edges we call \emph{skip edge} and \emph{backbone edge}, depending on \alg's behavior later on). 
The backbone edge and skip edge are incident to vertices of other gadgets.
The return edge is incident to a new vertex that is incident to two edges of weight $0$. 
One of those two edges of weight $0$ is incident to $u$. 
The other edge of weight $0$ is incident to a new vertex, which we call \emph{tail vertex}~$v_\text{t}$.
\end{NBG}

\begin{Def}[Exploration]\label{definition: Exploration}
\emph{Exploring} a block means visiting every vertex that is included inside of the block.
\end{Def}

\begin{restatable}{Lem}{LemmaCostEasy}\label{lemma: Costs Easy}
\alg incurs a cost of at least $2x-1$ inside of a normal block until every vertex of the block is explored and the last explored vertex is one of the last three vertices on the tail side.
\end{restatable}
\newcommand{\LemmaCostEasyProof}{
\begin{proof}
\alg starts at vertex $v_{\text{start}}$ and incurs a cost of at least $x-1$ until it reaches the head vertex.
At this point only the three vertices on the tail side are left to explore.
\alg can either backtrack to the first unvisited vertex on the other side of the path or take one of the edges of weight $x$. 
Both cases add a cost of $x$. 
So, the total cost of \alg for exploring $B$ is at least $2x-1$ since the remaining unexplored vertices of the block are adjacent to edges of weight $0$.
\end{proof}
}
\full{\LemmaCostEasyProof}

\subsection{Special Blocks}\label{subsection: Kinds of Blocks}

The idea now is to connect multiple normal blocks to each other and form a cycle of connected gadgets. 
We will construct our graph $G_\text{simple}$ in a way, such that normal blocks are always entered from a block connected to their tail vertex. 
The gadget containing the origin, called \emph{origin block}, is different in the sense that, initially, it is surrounded by gadgets that are not yet visited by \alg. 
Therefore, we have to construct it in a way that it makes no difference through which side \alg chooses to leave it. 
Likewise, the last block that is explored by \alg is different in the sense that it is surrounded by blocks that are already visited. 
Therefore, it has\shorten{ to have} some special structure to be able to act as link connecting both ends of the path of blocks to a cycle.
We call this special gadget \emph{closing block} since it closes the cycle of blocks.

\newcommand{\figureOriginBlock}{
\begin{figure}\conf{[p]}
\begin{center}
\begin{tikzpicture}[scale=0.28,font=\footnotesize,decoration=brace]

\draw[line width=1pt,black,-] (-30,0)--(-10,0)--(-10,10)--(-30,10)--(-30,0);

\draw (-28.8,4.8) node[below] {$v_{\text{h}}$};
\draw (-28,5) node[above] {$1$};
\draw (-26,5) node[above] {$1$};
\draw (-21,5) node[above] {$1$};
\draw (-20,5) node[below] {$v_o$};
\draw (-19,5) node[above] {$1$};
\draw (-14,5) node[above] {$1$};
\draw (-12,5) node[above] {$1$};
\draw (-20,0) node[below] {$O$};
\draw (-11.2,4.8) node[below] {$v_{\text{h}}$};

\draw (-29,0) node[below] {head};
\draw (-11,0) node[below] {head};

\draw (-10,8) node[right] {\scriptsize skip,};
\draw (-10,7) node[right] {\scriptsize $x$};
\draw (-10,3) node[right] {\scriptsize $x,$};
\draw (-10,2) node[right] {\scriptsize backbone};

\draw (-30,8) node[left] {\scriptsize skip,};
\draw (-30,7) node[left] {\scriptsize $x$};
\draw (-30,2) node[left] {\scriptsize backbone};
\draw (-30,3) node[left] {\scriptsize $x,$};

\filldraw[black] (-29,5) circle (0.25);
\filldraw[black] (-27,5) circle (0.25);
\filldraw[black] (-25,5) circle (0.25);
\filldraw[black] (-22,5) circle (0.25);
\filldraw[black] (-20,5) circle (0.25);
\filldraw[black] (-18,5) circle (0.25);
\filldraw[black] (-15,5) circle (0.25);
\filldraw[black] (-13,5) circle (0.25);
\filldraw[black] (-11,5) circle (0.25);

\draw (-23.5,5) node {\LARGE ...};
\draw (-16.5,5) node {\LARGE ...};

\draw[line width=1.5pt, black, -] (-29,5)--(-24.5,5);
\draw[line width=1.5pt, black, -] (-22.5,5)--(-17.5,5);
\draw[line width=1.5pt, black, -] (-15.5,5)--(-11,5);

\draw[line width=1.5pt, black, -] (-11,5) -- (-4,8);
\draw[line width=1.5pt, black, -] (-11,5) -- (-4,2);
\draw[line width=1.5pt, black, -] (-29,5) -- (-36,8);
\draw[line width=1.5pt, black, -] (-29,5) -- (-36,2);

\draw[line width=1pt,black,-] (0,0)--(20,0)--(20,10)--(0,10)--(0,0);

\draw (1.2,5) node[below] {$v_{\text{t}}$};
\draw (2,5) node[above] {$0$};
\draw (4,5) node[above] {$0$};
\draw (6,5) node[above] {$1$};
\draw (10,5) node[below] {$v_\text{start}$};
\draw (15.5,4.5) node[below] {$v_\text{ph}$};
\draw (17,5) node[below] {$v_\text{ps}$};
\draw (14,5) node[above] {$1$};
\draw (16,5) node[above] {$0$};
\draw (18,5) node[above] {$0$};
\draw (18.8,5) node[below] {$v_{\text{t}}$};

\draw (9,2) node[below] {return, $x$};

\draw (10,0) node[below] {$C$};

\draw (1,0) node[below] {tail};
\draw (19,0) node[below] {tail};

\foreach \x in {1,3,5,7,10,13,15,17,19}{
	\filldraw[black] (\x,5) circle (0.25);
}

\draw (8.5,5) node {\LARGE ...};
\draw (11.5,5) node {\LARGE ...};

\draw[line width=1.5pt, black, -] (1,5)--(7.5,5);
\draw[line width=1.5pt, black, -] (9.5,5)--(10.5,5);
\draw[line width=1.5pt, black, -] (12.5,5)--(19,5);

\draw[line width=1.5pt, black, -] (15,5) to[bend left] (12,2);
\draw[line width=1.5pt, black, -] (12,2)--(6,2);
\draw[line width=1.5pt, black, -] (3,5) to[bend right] (6,2);

\draw[decorate, yshift=2ex, thick]  (-29,6.75) -- node[above=0.4ex] {$x+1$ vertices}  (-11,6.75);
\draw[decorate, yshift=2ex, thick]  (5,6.75) -- node[above=0.4ex] {$x+1$ vertices}  (15,6.75);
\end{tikzpicture}
\end{center}
\captionshift
\caption{An origin block (left) and a closing block (right).}\label{figure: Origin Block}
\end{figure}
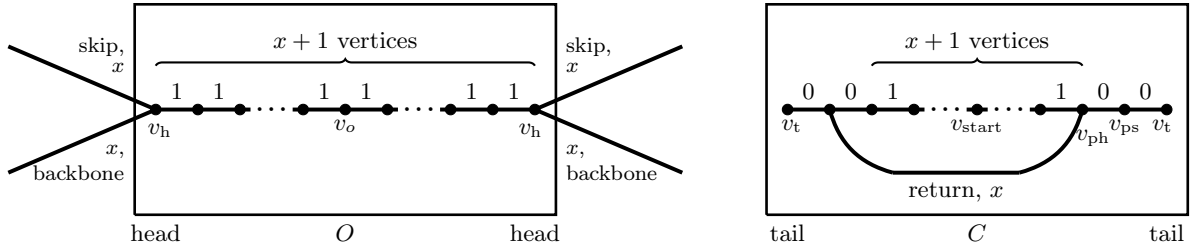
}
\full{\figureOriginBlock}

\begin{OBG}\label{definition: Origin Block}
An \textit{origin block} $O$ (see \Cref{figure: Origin Block}) is a subgraph of $G_\text{simple}$ that is a path of $x+1$ vertices. 
It contains the origin $v_o$, which has two incident edges of weight $1$. 
The basic construction is the same as a normal block, however, there are no edges of weight $0$ and we call both end vertices of the path \emph{head vertices}. 
Furthermore, there is no return edge. 
Instead, both head vertices have an incident skip edge and a backbone edge.
\end{OBG}

\begin{FBG}\label{definition: Final Block}
A \textit{closing block} $C$ (see \Cref{figure: Origin Block}) is a subgraph of $G_\text{simple}$ that is a path of $x+5$ vertices. 
The basic construction is the same as a normal block, however, there are two edges of weight~$0$ on both sides of the path and we call both end vertices of the path \emph{tail vertices}. 
Furthermore, there is no skip edge. 
We call the unique vertex of $C$ that is incident to one edge of weight $0$, one edge of weight $1$ and the return edge the \emph{pseudo head vertex}~$v_{\text{ph}}$.
Furthermore, we call the unique vertex of $C$ that is connected to the pseudo head vertex via an edge of weight $0$ the \emph{pseudo start vertex}~$v_{\text{ps}}$.
\end{FBG}

\begin{Obs}\label{lemma: Compatibility}
We can always connect the head side of a block to the tail side of another block by connecting a backbone edge to a tail vertex and a skip edge to a start vertex $v_\text{start}$. This can be done independently of the types of the two blocks.
\end{Obs}

\newcommand{\figureCompatibility}{
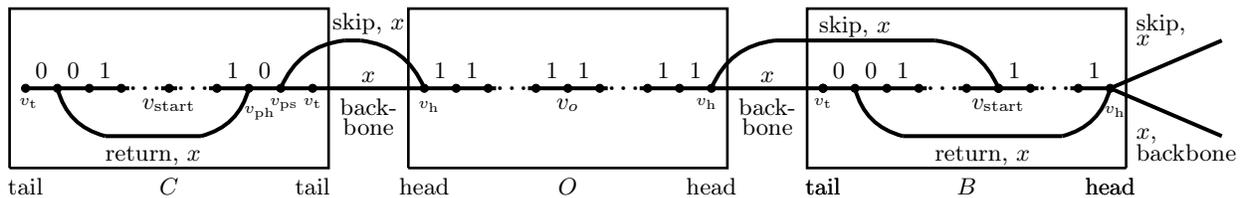
\begin{figure}\conf{[p]}
\begin{center}
\begin{tikzpicture}[scale=0.21205,font=\footnotesize]
\draw[line width=1pt,black,-] (-35,0)--(-15,0)--(-15,10)--(-35,10)--(-35,0);
\draw[line width=1pt,black,-] (-10,0)--(10,0)--(10,10)--(-10,10)--(-10,0);
\draw[line width=1pt,black,-] (35,0)--(15,0)--(15,10)--(35,10)--(35,0);

\draw (-8.7,4.8) node[below] {\tiny $v_{\text{h}}$};
\draw (-8,5) node[above] {$1$};
\draw (-6,5) node[above] {$1$};
\draw (-1,5) node[above] {$1$};
\draw (0,5) node[below] {$v_o$};
\draw (1,5) node[above] {$1$};
\draw (6,5) node[above] {$1$};
\draw (8,5) node[above] {$1$};
\draw (8.7,5) node[below] {\tiny $v_{\text{h}}$};

\draw (12.5,4.75) node[above] {$x$};
\draw (12.5,5) node[below] {back-};
\draw (12.5,3.75) node[below] {bone};

\draw (-12.5,7.8) node[above] {skip, $x$};

\draw (-12.5,4.75) node[above] {$x$};
\draw (-12.5,5) node[below] {back-};
\draw (-12.5,3.75) node[below] {bone};

\draw (18,7.6) node[above] {skip, $x$};

\draw (0,0) node[below] {$O$};
\draw (25,0) node[below] {$B$};

\draw (34,0) node[below] {head};
\draw (9,0) node[below] {head};
\draw (16,0) node[below] {tail};
\draw (-9,0) node[below] {head};

\filldraw[black] (-9,5) circle (0.25);
\filldraw[black] (-7,5) circle (0.25);
\filldraw[black] (-5,5) circle (0.25);
\filldraw[black] (-2,5) circle (0.25);
\filldraw[black] (0,5) circle (0.25);
\filldraw[black] (2,5) circle (0.25);
\filldraw[black] (5,5) circle (0.25);
\filldraw[black] (7,5) circle (0.25);
\filldraw[black] (9,5) circle (0.25);

\filldraw[black] (16,5) circle (0.25);
\filldraw[black] (27,5) circle (0.25);

\draw (-3.5,5) node {\LARGE ...};
\draw (3.5,5) node {\LARGE ...};

\draw[line width=1.5pt, black, -] (-16,5)--(-4.5,5);
\draw[line width=1.5pt, black, -] (-2.5,5)--(2.5,5);
\draw[line width=1.5pt, black, -] (4.5,5)--(16,5);

\draw[line width=1.5pt, black, -] (9,5) to[bend left] (13,8);
\draw[line width=1.5pt, black, -] (13,8)--(23,8);
\draw[line width=1.5pt, black, -] (27,5) to[bend right] (23,8);

\draw[line width=1.5pt, black, -] (-9,5) to[bend right] (-13,8);
\draw[line width=1.5pt, black, -] (-13,8)--(-14,8);
\draw[line width=1.5pt, black, -] (-14,8) to[bend right] (-18,5);

\draw (16,5) node[below] {\tiny $v_{\text{t}}$};
\draw (17,5) node[above] {$0$};
\draw (19,5) node[above] {$0$};
\draw (28,5) node[above] {$1$};
\draw (27,5) node[below] {$v_{\text{start}}$};
\draw (21,5) node[above] {$1$};
\draw (33,5) node[above] {$1$};
\draw (34.3,4.5) node[below] {\tiny $v_{\text{h}}$};

\draw (35,9) node[right] {skip,};
\draw (35,8) node[right] {$x$};
\draw (35,1) node[right] {backbone};
\draw (35,2) node[right] {$x$,};

\draw (26,2) node[below] {return, $x$};

\draw (16,0) node[below] {tail};
\draw (34,0) node[below] {head};

\foreach \x in{16,18,20,22,25,27,29,32,34}{
	\filldraw[black] (\x,5) circle (0.25);
}

\draw (23.5,5) node {\LARGE ...};
\draw (30.5,5) node {\LARGE ...};

\draw[line width=1.5pt, black, -] (16,5)--(22.5,5);
\draw[line width=1.5pt, black, -] (24.5,5)--(29.5,5);
\draw[line width=1.5pt, black, -] (31.5,5)--(34,5);

\draw[line width=1.5pt, black, -] (34,5) -- (41,8);
\draw[line width=1.5pt, black, -] (34,5) -- (41,2);

\draw[line width=1.5pt, black, -] (34,5) to[bend left] (31,2);
\draw[line width=1.5pt, black, -] (31,2)--(21,2);
\draw[line width=1.5pt, black, -] (18,5) to[bend right] (21,2);

\draw (1.2-35,5) node[below] {\tiny $v_{\text{t}}$};
\draw (2-35,5) node[above] {$0$};
\draw (4-35,5) node[above] {$0$};
\draw (6-35,5) node[above] {$1$};
\draw (10-35,5) node[below] {$v_\text{start}$};
\draw (15.7-35,4.5) node[below] {\tiny $v_\text{ph}$};
\draw (17.2-35,5) node[below] {\tiny $v_\text{ps}$};
\draw (14-35,5) node[above] {$1$};
\draw (16-35,5) node[above] {$0$};
\draw (19-35,5) node[below] {\tiny $v_{\text{t}}$};

\draw (9-35,2) node[below] {return, $x$};

\draw (10-35,0) node[below] {$C$};

\draw (1-35,0) node[below] {tail};
\draw (19-35,0) node[below] {tail};

\foreach \x in {1,3,5,7,10,13,15,17,19}{
	\filldraw[black] (\x-35,5) circle (0.25);
}

\draw (8.5-35,5) node {\LARGE ...};
\draw (11.5-35,5) node {\LARGE ...};

\draw[line width=1.5pt, black, -] (1-35,5)--(7.5-35,5);
\draw[line width=1.5pt, black, -] (9.5-35,5)--(10.5-35,5);
\draw[line width=1.5pt, black, -] (12.5-35,5)--(19-35,5);

\draw[line width=1.5pt, black, -] (15-35,5) to[bend left] (12-35,2);
\draw[line width=1.5pt, black, -] (12-35,2)--(6-35,2);
\draw[line width=1.5pt, black, -] (3-35,5) to[bend right] (6-35,2);
\end{tikzpicture}
\end{center}
\captionshift
\caption{A closing block connected to an origin block connected to a normal block. Note that the scenario drawn in this figure occurs if the agent never backtracks and only explores blocks to the right of the origin block. In this case the closing block is first entered from the left via the $x$-th explored normal block.}\label{figure: Compatibility}
\end{figure}
}
\full{\figureCompatibility}

See \Cref{figure: Compatibility} for an example of a closing block connected to an origin block connected to a normal block. 
Note that in case of a closing block, we connect the pseudo start vertex and its adjacent tail vertex to the same head side (in case of \Cref{figure: Compatibility} one head side of an origin block).
Both special blocks as well as normal blocks can be explored optimally with cost $x$ by omitting skip and return edges and only taking backbone edges between blocks.

\begin{restatable}{Lem}{LemmaCostEasyOpt}\label{lemma: Costs Easy Opt}
Let $B$ be any block. 
Assume $B$ has backbone edges incident to all tail and head vertices (i.e., by construction, the block contains exactly two vertices with incident backbone edges; one on each side) and assume $B$ is entered via a backbone edge. Exploring the whole block and reaching the vertex incident to the backbone edge on the other side incurs costs of $x$.
\end{restatable}

\subsection{Analysis}\label{subsection: The Strategy 2}

We start by formally defining the \emph{examination of a block} and then describe the macroscopic design of the graph $G_{\text{simple}}$, i.e., how the blocks are connected to each other.

\begin{Def}[Examination]\label{definition: Examination}
A normal block is \emph{examined} once its head vertex has been visited.
A closing block is \emph{examined} once its pseudo-head vertex has been visited.
An origin block is \emph{examined} once \alg has left it for the first time.
\end{Def}

\begin{GTwo}
\alg starts at the origin $v_o$ inside an origin block.
We let \alg explore the block until it reaches a head vertex leading outside of the origin block.
In order to continue the exploration, at some point, \alg takes one of the two edges of weight $x$ at a head vertex of the origin block.
Since \alg cannot distinguish between the endpoints of the edges, we let the edge chosen by \alg be the skip edge leading to a vertex $v_\text{start}$ of a normal block.
If \alg chooses to backtrack and leave the origin block through the other side of it, it enters a normal block via a skip edge in the same way.
Note it is possible to connect normal blocks to both sides of an origin block by \Cref{lemma: Compatibility}.
If \alg leaves a normal block $B$ via the skip edge it already used to enter $B$ before it has explored the head vertex of $B$ and renters $B$ at the tail side via the backbone edge, we connect one end of the explored path in $B$ to the vertex inside of $B$ that is adjacent to the tail vertex. 
Every time \alg is at a head vertex of a normal block and takes one of the two remaining edges of weight $x$ (after we have let it take the return edge the first time), we let it be the skip edge. 
By construction, \alg then arrives at the (pseudo-)start vertex of an unexamined block.
Every new block examined by \alg is a normal block until \alg has examined $x+2$ blocks (including the origin block).
We let the $(x+3)$-rd block examined by \alg be a closing block connecting both ends of the path of blocks (see \Cref{figure: Simple Circle Overview}).
Note that both sides of a closing block can be connected to a normal block or an origin block by \Cref{lemma: Compatibility}.
\end{GTwo}

Note that $G_\text{simple}$ is a planar graph: 
If we exclude the skip and return edges, $G_\text{simple}$ is just a cycle.
If we embed this cycle on a plane, we can add the skip edges on the outer side of the cycle and the return edges on the inner side of the cycle.
The planarity of $G_\text{simple}$ then follows from the fact that skip edges never intersect with each other and return edges never intersect with each other.

It remains to examine the cost of \alg for exploring $G_\text{simple}$. In \Cref{lemma: Costs Easy}, we have seen that \alg incurs costs of at least $2x-1$ for exploring a block if it stays inside the block. Thus, it remains to make sure that \alg cannot reduce its costs by leaving the block prematurely. 

\begin{restatable}{Lem}{LemmaCostPerBlock}\label{lemma: Costs per Block}
Suppose $B$ is a normal block of $G_\text{simple}$ that has been entered by \alg via the skip edge of a block next to the tail side of~$B$ for the first time. 
Then \alg incurs a cost of at least $4x-1$ inside $B$ until it has explored $B$ and has entered the unexplored block adjacent to the head side of~$B$.
\end{restatable}
\newcommand{\LemmaCostPerBlockProof}{
\begin{proof}
Assume \alg leaves~$B$ before exploring it and instead goes back through the skip edge it first arrived by. 
This incurs a cost of~$x$.
For exploring~$B$, \alg has to reenter~$B$ either by walking the skip edge or by walking the backbone edge.
Both options incur a cost of~$x$ by \Cref{lemma: Costs Easy Opt}.
Visiting every vertex of~$B$ then again incurs a cost of at least $x$. 
Finally, \alg has to to take one of the edges of weight $x$ incident to the head vertex of $B$ to enter the unexplored block adjacent to the head side of $B$.
Thus, in total, \alg has a cost of at least~$4x$, already exceeding the claimed cost of $4x-1$.

Therefore, we may assume that \alg stays inside $B$ and explores $B$. 
According to \Cref{lemma: Costs Easy}, \alg incurs a cost of at least $2x-1$ for exploring $B$ and the vertex of $B$ explored last is one of the three vertices on the tail side of $B$.
To enter the unexplored block adjacent to the head side of $B$, \alg needs backtracks to the head vertex of $B$ for costs of $x$. 
Then, \alg has to take one of the edges of weight $x$ incident to the head vertex of $B$ to enter the unexplored block adjacent to the head side of $B$.
This results in a total cost of at least $4x-1$.
\end{proof}
}
\full{\LemmaCostPerBlockProof}

\newcommand{\TheoremLowerBoundTwoProof}{
\begin{proof}[Proof of \Cref{theorem: Lower Bound 2}]
Let $\varepsilon>0$. 
$G_\text{simple}$ consists of~$x+3$ blocks.
We can apply \Cref{lemma: Costs per Block} for at least~$x$ of the~$x+1$ normal blocks, accumulating a cost of at least $4x^2-x$. 
Note that \Cref{lemma: Costs per Block} might not be applicable for one normal block that is connected to the closing block since the closing block can be explored from two sides.
Thus, we have
\begin{equation*}
\alg(G_{\text{simple}})\geq 4x^2-x.
\end{equation*}
\opt on the other hand can just traverse every block in one direction and can transit via backbone edges, moving a distance of~$x$ per block (\Cref{lemma: Costs per Block}) and~$x$ per backbone edge.
Thus
\begin{equation*}
\opt(G_{\text{simple}})=(x+3)x+(x+3)x=2x^2+6x.
\end{equation*}
Therefore, this construction gives a lower bound for the competitive ratio of
\begin{equation*}
\frac{\alg(G_{\text{simple}})}{\opt(G_{\text{simple}})}\geq\frac{4x^2-x}{2x^2+6x}\overset{x\rightarrow\infty}{\longrightarrow} 2>2-\varepsilon.
\end{equation*}
\end{proof}
}
\full{
\TheoremLowerBoundTwoProof
}

\section{Recursive Construction: Lower Bound of \texorpdfstring{$3$}{3}}\label{section: Improved Recursive Construction}\label{section: Improved Lower Bound of 3}

We refine the approach of \Cref{section: Basic Construction}:
The macroscopic cycle structure of $G_\text{simple}$ remains the same as before, but we introduce a recursive construction inside the blocks, i.e., the blocks now consist of blocks instead of vertices.
Whenever we talk about a block $A$ that is inside a block~$B$, we say that~$A$ is a subblock of~$B$.
Furthermore, we introduce a constant $y\in\{0,\dots,\lfloor\frac{x}{2}\rfloor\}$ and construct every normal block (except the ones on the highest level) to have at least~$y$ subblocks between the origin subblock and last subblock on the head side (see \Cref{figure: Three Levels}).
This constant~$y$ ensures that there are at least $y$ subblocks to traverse between two skip edges, forcing the agent to accumulate additional costs in case of backtracking.

We call this refined graph $G_{\text{rec}}$ since it contains a recursive structure. 
In this graph, \alg accumulates at least $3-\frac{2}{N+2}$ times the cost the optimum accumulates, where $N$ is the number of recursive levels. 
This improves our lower bound to $3$.
In particular, we will show the following.

\begin{restatable}{The}{TheoremLowerBoundThree}\label{theorem: Lower Bound 3}
For every $\varepsilon>0$ there is a graph $G_\text{rec}$ such that $\alg(G_\text{rec})\ge (3-\varepsilon)\cdot\opt(G_\text{rec})$.
\end{restatable}

We now formalize the idea above.
In general, for $i>0$, the block of level~$i$ contains subblocks of level~$i-1$. 
The edges inside of the block of level~$i$ are the backbone/skip edges of subblocks of level~$i-1$. 
Blocks of level 0 contain vertices instead of subblocks and are similarly structured as the blocks of \Cref{section: Basic Construction}.

\newcommand{\figureThreeLevels}{
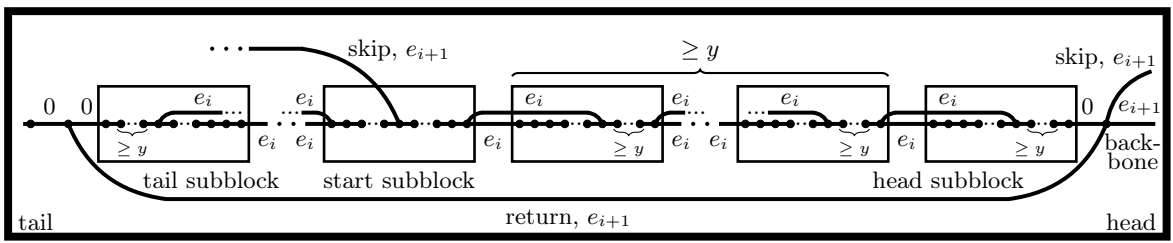
\begin{figure}\conf{[p]}
\begin{center}
\begin{tikzpicture}[scale=0.1,font=\footnotesize,decoration=brace]

\draw[line width=3pt] (-77,-10) rectangle (77,20);

\draw[line width=1pt] (-65,0) rectangle (-45,10);
\draw[line width=1pt] (-35,0) rectangle (-15,10);
\draw[line width=1pt] (-10,0) rectangle (10,10);
\draw[line width=1pt] (40,0) rectangle (20,10);
\draw[line width=1pt] (65,0) rectangle (45,10);

\draw (-71.5,5) node[above] {$0$};
\draw (-66.5,5) node[above] {$0$};
\draw (-42.5,4.75) node[below] {$\eX{i}$};
\draw (-37.5,4.75) node[below] {$\eX{i}$};
\draw (-37.5,6) node[above] {$\eX{i}$};
\draw (42.5,4.75) node[below] {$\eX{i}$};
\draw (73.5,4.75) node[above] {$\eX{i+1}$};
\draw (72.5,5) node[below] {back-};
\draw (72.5,2.5) node[below] {bone};
\draw (69,11) node[above] {skip, $\eX{i+1}$};
\draw (66.5,5) node[above] {$0$};
\draw (-2.5,-5) node[below] {return, $\eX{i+1}$};
\draw (-50,0) node[below] {tail subblock};
\draw (48,0) node[below] {head subblock};
\draw (-25,0) node[below] {start subblock};

\draw (-51,6) node[above] {$\eX{i}$};
\draw (48,6) node[above] {$\eX{i}$};
\draw (27,6) node[above] {$\eX{i}$};
\draw (-7,6) node[above] {$\eX{i}$};

\foreach \x in {-74,-69,-64,-62,-59,-57,-55,-52,-50,-48,-46, 64, 62, 59, 57, 55, 52, 50, 48, 46,69}{
	\filldraw[black] (\x,5) circle (0.5);
}

\draw[line width=1.5pt, black, -] (-57,5) to[bend left] (-55,6.5);
\draw[line width=1.5pt, black, -] (-55.1,6.5)--(-48.5,6.5);
\draw (-47,6.5) node {\footnotesize ...};

\draw[line width=1.5pt, black, -] (-34,5) to[bend right] (-36,6.5);
\draw[line width=1.5pt, black, -] (-35.9,6.5)--(-38,6.5);
\draw (-39.5,6.5) node {\footnotesize ...};

\draw[line width=1.5pt, black, -] (-16,5) to[bend left] (-14,6.5);
\draw[line width=1.5pt, black, -] (-14.1,6.5)--(0.1,6.5);
\draw[line width=1.5pt, black, -] (2,5) to[bend right] (0,6.5);

\draw[line width=1.5pt, black, -] (9,5) to[bend left] (11,6.5);
\draw[line width=1.5pt, black, -] (10.9,6.5)--(13,6.5);
\draw (14.5,6.5) node {\footnotesize ...};

\draw[line width=1.5pt, black, -] (39,5) to[bend left] (41,6.5);
\draw[line width=1.5pt, black, -] (40.9,6.5)--(55.1,6.5);
\draw[line width=1.5pt, black, -] (57,5) to[bend right] (55,6.5);

\draw[line width=1.5pt, black, -] (32,5) to[bend right] (30,6.5);
\draw[line width=1.5pt, black, -] (30.1,6.5)--(24,6.5);
\draw (22.5,6.5) node {\footnotesize ...};

\draw[line width=1.5pt, black, -] (-75,5)--(-61.5,5);
\draw[line width=1.5pt, black, -] (-59.5,5)--(-54.5,5);
\draw[line width=1.5pt, black, -] (-52.5,5)--(-42.5,5);

\draw[line width=1.5pt, black, -] (-37.5,5)--(-29.5,5);
\draw[line width=1.5pt, black, -] (-27.5,5)--(-22.5,5);
\draw[line width=1.5pt, black, -] (-20.5,5)--(-2.5,5);
\draw[line width=1.5pt, black, -] (-0.5,5)--(4.5,5);
\draw[line width=1.5pt, black, -] (6.5,5)--(12.5,5);

\draw[line width=1.5pt, black, -] (17.5,5)--(27.5,5);
\draw[line width=1.5pt, black, -] (29.5,5)--(34.5,5);
\draw[line width=1.5pt, black, -] (36.5,5)--(52.5,5);
\draw[line width=1.5pt, black, -] (54.5,5)--(59.5,5);
\draw[line width=1.5pt, black, -] (61.5,5)--(75.5,5);

\draw (-60.5,5) node {\footnotesize ..};
\draw (-53.5,5) node {\footnotesize ..};
\draw (-28.5,5) node {\footnotesize ..};
\draw (-21.5,5) node {\footnotesize ..};
\draw (-1.5,5) node {\footnotesize ..};
\draw (5.5,5) node {\footnotesize ..};
\draw (28.5,5) node {\footnotesize ..};
\draw (35.5,5) node {\footnotesize ..};
\draw (53.5,5) node {\footnotesize ..};
\draw (60.5,5) node {\footnotesize ..};

\draw (-40,5) node {\huge ..};
\draw (15,5) node {\huge ..};

\draw[line width=1.5pt, black, -] (69,5) to[bend left] (56,-5);
\draw[line width=1.5pt, black, -] (-56.2,-5) -- (56.2,-5);
\draw[line width=1.5pt, black, -] (-56,-5) to[bend left] (-69,5);

\draw[line width=1.5pt, black, -] (69,5) to[bend left] (75,12);

\draw (-25,17.5) node[below] {skip, $\eX{i+1}$};

\draw (-12.5,4.75) node[below] {$\eX{i}$};
\draw (12.5,4.75) node[below] {$\eX{i}$};
\draw (17.5,4.75) node[below] {$\eX{i}$};
\draw (12.5,6) node[above] {$\eX{i}$};

\draw (-77,-8) node[right] {tail};
\draw (77,-8) node[left] {head};

\foreach \x in {-34,-32,-30,-27,-25,-23,-20,-18,-16}{
	\filldraw[black] (\x,5) circle (0.5);
}

\foreach \x in {-9,-7,-5,-3,0,2,4,7,9}{
	\filldraw[black] (\x,5) circle (0.5);
}

\foreach \x in {21,23,25,27,30,32,34,37,39}{
	\filldraw[black] (\x,5) circle (0.5);
}

\draw[line width=1.5pt, black, -] (-25,5) to[bend right] (-38,15);
\draw[line width=1.5pt, black, -] (-37.8,15)--(-45,15);
\draw (-48,15) node {\huge ...};

\draw[decorate, yshift=2ex, thick]  (-10,11) -- node[above=0.4ex] {$\ge y$}  (40,11);
\draw[decorate, yshift=2ex]  (7.5,4) -- node[below=0.2ex] {\tiny $\ge y$}  (3.5,4);
\draw[decorate, yshift=2ex]  (37.5,4) -- node[below=0.2ex] {\tiny $\ge y$}  (33.5,4);
\draw[decorate, yshift=2ex]  (-58.5,4) -- node[below=0.2ex] {\tiny $\ge y$}  (-62.5,4);
\draw[decorate, yshift=2ex]  (62.5,4) -- node[below=0.2ex,xshift=-1ex] {\tiny $\ge y$}  (58.5,4);
\end{tikzpicture}
\end{center}
\captionshift
\caption{Example of a block of level $i+1$. Instead of vertices it consists of blocks of level $i$.
The basic design of the blocks from \Cref{section: Basic Construction} remains the same, i.e., there is a skip edge leading to some block in the middle of the path of blocks, called origin subblock.
As before, there is a return edge going from the \emph{head vertex} to a vertex at the tail side. 
Additionally, there are at least $y$ subblocks between the origin subblock and the subblock connected to the head vertex.
The dots inside the blocks of level $i$ represent blocks of level $i-1$.
The edge weights will be defined later.}\label{figure: Three Levels}
\end{figure}
}
\full{\figureThreeLevels}

\Cref{figure: Three Levels} shows the interaction of three levels. 
It shows a block of level $i+1$ consisting of subblocks of level $i$.
These subblocks replace the vertices of the blocks from \Cref{section: Basic Construction}.
Note that the block that is marked as \emph{start subblock} as well as the rightmost and the leftmost block, which are marked as \emph{head subblock} and \emph{tail subblock} in \Cref{figure: Three Levels}, differ from the other blocks and thus have a special construction. 
The block that is marked as \emph{start subblock} shares similarity with the origin block of \Cref{section: Basic Construction}, i.e., it is the first block of level $i$ in \Cref{figure: Three Levels} that is entered by \alg and is surrounded by two unvisited blocks. 
Fittingly, we will call these type of blocks \emph{origin blocks of level $i$}. 
The blocks of level $i$ , which are marked as \emph{head subblock} and \emph{tail subblock} in \Cref{figure: Three Levels} are special in that they have no skip edges. 
We will call these type of blocks \emph{final blocks of level $i$}. 
All remaining blocks are called \emph{normal blocks of level $i$}.
We give a detailed description of all blocks in the next subsections.
By $\eX{i}$ we denote the weight of a backbone edge of level $i$ and for now postpone the exact definition.
The skip edges and return edges of a block of level $i$ also have weight~$\eX{i}$.

\subsection{Recursion Basis}\label{subsection: Level 0: The Lowest level}

The blocks on level $0$ are very similar to the blocks of \Cref{section: Basic Construction}. 
For convenience, we define $e_{-1}\coloneqq 1$ to be the weight of an edge connecting two vertices in a block of level $0$. 
Furthermore, we define $\eX{0}\coloneqq x$ to be the weight of skip/return/backbone edges of level $0$.

\newcommand{\figureNormalBlockO}{
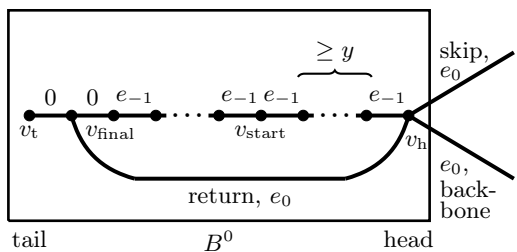
\begin{figure}\conf{[p]}
\begin{center}
\begin{tikzpicture}[scale=0.28,font=\footnotesize,decoration=brace]
\draw[line width=1pt,black,-] (-30-10,0)--(-10-10,0)--(-10-10,10)--(-30-10,10)--(-30-10,0);

\draw (-29-10,5) node[below] {$v_{\text{t}}$};
\draw (-25.15-10,5) node[below] {$v_\text{final}$};
\draw (-28-10,5) node[above] {$0$};
\draw (-26-10,5) node[above] {$0$};
\draw (-17-10,5) node[above] {$\eX{-1}$};
\draw (-19-10,5) node[above] {$\eX{-1}$};
\draw (-18-10,5) node[below] {$v_{\text{start}}$};
\draw (-24-10,5) node[above] {$\eX{-1}$};
\draw (-12-10,5) node[above] {$\eX{-1}$};
\draw (-10.6-10,4.5) node[below] {$v_{\text{h}}$};

\draw (-10-10,8) node[right] {skip,};
\draw (-10-10,7) node[right] {$\eX{0}$};
\draw (-10-10,2.5) node[right] {$\eX{0}$,};
\draw (-10-10,1.5) node[right] {back-};
\draw (-10-10,0.5) node[right] {bone};

\draw (-19-10,2) node[below] {return, $\eX{0}$};

\draw (-20-10,0) node[below] {$B^0$};

\draw (-29-10,0) node[below] {tail};
\draw (-11-10,0) node[below] {head};

\foreach \x in{-29,-27,-25,-23,-20,-18,-16,-13,-11}{
	\filldraw[black] (\x-10,5) circle (0.25);
}

\draw (-21.5-10,5) node {\LARGE ...};
\draw (-14.5-10,5) node {\LARGE ...};

\draw[line width=1.5pt, black, -] (-29-10,5)--(-22.5-10,5);
\draw[line width=1.5pt, black, -] (-20.5-10,5)--(-15.5-10,5);
\draw[line width=1.5pt, black, -] (-13.5-10,5)--(-11-10,5);

\draw[line width=1.5pt, black, -] (-11-10,5) -- (-6-10,8);
\draw[line width=1.5pt, black, -] (-11-10,5) -- (-6-10,2);
\draw[line width=1.5pt, black, -] (-11-10,5) to[bend left] (-14-10,2);
\draw[line width=1.5pt, black, -] (-14-10,2)--(-24-10,2);
\draw[line width=1.5pt, black, -] (-27-10,5) to[bend right] (-24-10,2);
\draw[decorate, yshift=2ex, thick]  (-26.25,6.5) -- node[above=0.4ex] {$\ge y$}  (-22.75,6.5);

\end{tikzpicture}
\captionshift
\caption{A normal block of type $B^0$.}\label{figure: Normal Block 0}
\end{center}
\end{figure}
}
\full{\figureNormalBlockO}

\begin{NBGO}\label{strategy: Simple Block Level O}
A normal block of level $0$ (see \Cref{figure: Normal Block 0}), also called \textit{block of type~$B^0$}, is defined as follows.
Assume \alg starts at some vertex $v_{\text{start}}$ and has two incident edges of weight~$\eX{-1}$ to choose from.
Every time \alg chooses to traverse an edge of weight~$\eX{-1}$, it arrives at a new vertex that is incident to another edge of weight~$\eX{-1}$.
This way, \alg explores the graph and every new vertex it visits is incident to exactly one unexplored edge of weight~$\eX{-1}$. 
We stop this procedure once \alg has explored~$x+2$ vertices.
In the case that $y$ or more vertices lie on the path from the currently visited vertex to $v_{\text{start}}$ (excluding the currently visited vertex and $v_{\text{start}}$), we call the currently visited vertex \emph{head vertex} $v_\text{h}$ and call the first unvisited vertex to the other side of the path \emph{final vertex} $v_\text{final}$.
Otherwise, we call the currently visited vertex final vertex and call the first unvisited vertex to the other side of the path head vertex.
Note that in both cases there are at least $y\le\lfloor\frac{x}{2}\rfloor$ vertices on the path from the $v_{\text{h}}$ to $v_{\text{start}}$ (excluding $v_{\text{h}}$ and $v_{\text{start}}$).
The head vertex is incident to three edges of weight~$x$.
If \alg is at the head vertex and chooses to traverse an edge of weight~$x$, we call it a \textit{return edge} (the other two edges we call \emph{skip edge} and \emph{backbone edge}, depending on \alg's behavior later on). 
The backbone edge and skip edge are incident to vertices of other gadgets.
The return edge is incident to a new vertex that is incident to two edges of weight~$0$. 
One of those two edges of weight~$0$ is incident to the final vertex. 
The other edge of weight~$0$ is incident to a new vertex, which we call \emph{tail vertex} $v_\text{t}$.
\end{NBGO}

\begin{restatable}{Lem}{LemmaCostEasyLevelO}\label{lemma: Costs Easy Level O}
Let $B$ be a normal block of type $B^0$ that \alg entered via skip edge, i.e., the vertex of $B$ that is first visited by \alg is the start vertex.
\begin{enumerate}[topsep=0.5ex,itemsep=0.5ex,partopsep=0.5ex,parsep=0.5ex]
\item If the $(x+2)$-nd explored vertex in $B$ is the head vertex, then
\begin{itemize}[topsep=0.5ex,itemsep=0ex,partopsep=0.5ex,parsep=0.5ex]
\item \alg's cost inside $B$ until $B$ is fully explored is at least $2x+1$ and
\item \alg's last explored vertex of $B$ is one of the last three vertices on the tail side.
\end{itemize}
\item If the $(x+2)$-nd explored vertex in $B$ is the final vertex, then either
\begin{itemize}[topsep=0.5ex,itemsep=0ex,partopsep=0.5ex,parsep=0.5ex]
\item \alg's cost inside $B$ until $B$ is fully explored is at least $3x-y+1$ and
\item \alg's last explored vertex of $B$ is the head vertex
\end{itemize}
\vspace*{-1ex}
or
\vspace*{-1ex}
\begin{itemize}[topsep=0.5ex,itemsep=0ex,partopsep=0.5ex,parsep=0.5ex]
\item \alg's cost inside $B$ until $B$ is fully explored is at least $4x-y+1$ and
\item \alg's last explored vertex of $B$ is one of the last two vertices on the tail side.
\end{itemize}
\end{enumerate}
\end{restatable}
\newcommand{\LemmaCostEasyLevelOProof}{
\begin{proof}
Assume the $(x+2)$-nd vertex of $B$ explored by \alg is the head vertex.
Then there have to be at least $y$ vertices between the $(x+2)$-nd vertex explored by \alg and $v_{\text{start}}$.
The cheapest way for \alg be in this scenario is by walking straight in one direction and incurring costs of $x+1$ until it has explored $x+2$ vertices and reached the head vertex.
It remains to explore the last three vertices on the tail side. 
The head vertex has three incident edges of weight $\eX{0}$.
\alg can either backtrack to the first unvisited vertex on the tail side or take one of the edges of weight $\eX{0}=x$, which, by construction, will be the return edge.  
Both cases add a cost of at least $x$ (backtracking costs $x+2$), raising the total cost to at least $2x+1$.

Now assume the $(x+2)$-nd vertex explored by \alg is the final vertex.
Then there have to be at most $y-1$ vertices between $(x+2)$-nd vertex explored by \alg and $v_{\text{start}}$.
In other words, \alg has explored at least $x-y+1$ vertices on the other side of the vertex $v_{\text{start}}$.
Thus, \alg has incurred costs of at least $2(x-y+1)+(y-1)$ until until it has explored $x+2$ vertices.
It remains to explore the last two vertices on the tail side and the head vertex.

The last two vertices on the tail side can be explored for costs of $0$ since they are connected via $0$-weighted edges, i.e., \alg incurs no additional costs for exploring them.
Then \alg can either backtrack to the head side of the block or take the return edge of weight $\eX{0}=x$ to explore the head vertex. 
Both cases add a cost of at least $x$ (backtracking costs $x+2$), raising the total cost to at least $3x-y+1$ and the last explored vertex is the head vertex.

If \alg explores the head vertex before exploring both vertices on the tail side, it has to either backtrack to the head side of the block or take the return edge of weight $\eX{0}=x$ to explore the head vertex.
Both cases add a cost of at least $x$ (backtracking costs $x+2$).
Afterwards it has to backtrack yet again to the tail side to explore the remaining vertex there.
This adds again a cost of at least $x$, raising the total cost to at least $4x-y+1$.
\end{proof}
}
\full{\LemmaCostEasyLevelOProof}

\newcommand{\figureOriginBlockO}{
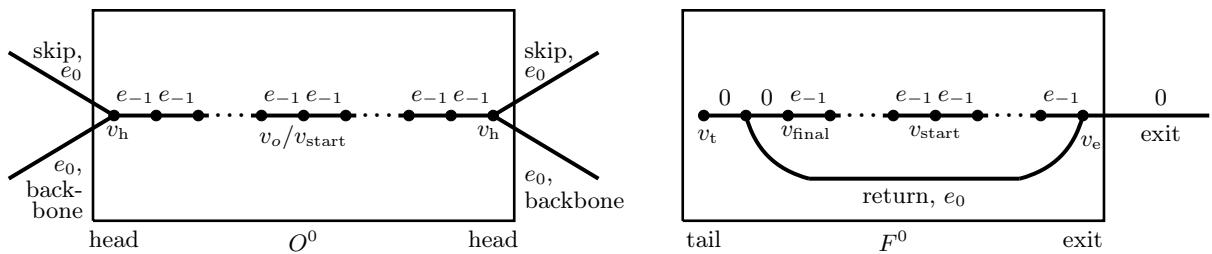
\begin{figure}\conf{[p]}
\begin{center}
\begin{tikzpicture}[scale=0.28,font=\footnotesize,decoration=brace]

\draw[line width=1pt,black,-] (-10,0)--(10,0)--(10,10)--(-10,10)--(-10,0);

\draw (-8.8,5) node[below] {$v_{\text{h}}$};
\draw (-8,5) node[above] {$\eX{-1}$};
\draw (-6,5) node[above] {$\eX{-1}$};
\draw (-1,5) node[above] {$\eX{-1}$};
\draw (0,5) node[below] {$v_o/v_\text{start}$};
\draw (1,5) node[above] {$\eX{-1}$};
\draw (6,5) node[above] {$\eX{-1}$};
\draw (8,5) node[above] {$\eX{-1}$};
\draw (0,0) node[below] {$O^0$};
\draw (8.8,5) node[below] {$v_{\text{h}}$};

\draw (-9,0) node[below] {head};
\draw (9,0) node[below] {head};

\filldraw[black] (-9,5) circle (0.25);
\filldraw[black] (-7,5) circle (0.25);
\filldraw[black] (-5,5) circle (0.25);
\filldraw[black] (-2,5) circle (0.25);
\filldraw[black] (0,5) circle (0.25);
\filldraw[black] (2,5) circle (0.25);
\filldraw[black] (5,5) circle (0.25);
\filldraw[black] (7,5) circle (0.25);
\filldraw[black] (9,5) circle (0.25);

\draw (-3.5,5) node {\LARGE ...};
\draw (3.5,5) node {\LARGE ...};

\draw[line width=1.5pt, black, -] (-9,5)--(-4.5,5);
\draw[line width=1.5pt, black, -] (-2.5,5)--(2.5,5);
\draw[line width=1.5pt, black, -] (4.5,5)--(9,5);

\draw[line width=1.5pt, black, -] (9,5) -- (14,8);
\draw[line width=1.5pt, black, -] (9,5) -- (14,2);
\draw[line width=1.5pt, black, -] (-9,5) -- (-14,8);
\draw[line width=1.5pt, black, -] (-9,5) -- (-14,2);

\draw (-10,8) node[left] {skip,};
\draw (-10,7) node[left] {$\eX{0}$};
\draw (-10,2.5) node[left] {$\eX{0}$,};
\draw (-10,1.5) node[left] {back-};
\draw (-10,0.5) node[left] {bone};

\draw (10,8) node[right] {skip,};
\draw (10,7) node[right] {$\eX{0}$};
\draw (10,2) node[right] {$\eX{0}$,};
\draw (10,1) node[right] {backbone};


\draw[line width=1pt,black,-] (-30+48,0)--(-10+48,0)--(-10+48,10)--(-30+48,10)--(-30+48,0);
\draw (-28.8+48,4.8) node[below] {$v_{\text{t}}$};
\draw (-24.15+48,5) node[below] {$v_\text{final}$};
\draw (-28+48,5) node[above] {$0$};
\draw (-26+48,5) node[above] {$0$};
\draw (-19+48,5) node[above] {$\eX{-1}$};
\draw (-18+48,5) node[below] {$v_\text{start}$};
\draw (-17+48,5) node[above] {$\eX{-1}$};
\draw (-24+48,5) node[above] {$\eX{-1}$};
\draw (-12+48,5) node[above] {$\eX{-1}$};
\draw (-10.6+48,4.5) node[below] {$v_{\text{e}}$};
\draw (-7.3+48,5) node[above] {$0$};
\draw (-7.3+48,5) node[below] {exit};


\draw (-19+48,2) node[below] {return, $\eX{0}$};

\foreach \x in{-29,-27,-25,-23,-20,-18,-16,-13,-11}{
	\filldraw[black] (\x+48,5) circle (0.25);
}

\draw (-20+48,0) node[below] {$F^0$};

\draw (-29+48,0) node[below] {tail};
\draw (-11+48,0) node[below] {exit};

\draw (-21.5+48,5) node {\LARGE ...};
\draw (-14.5+48,5) node {\LARGE ...};

\draw[line width=1.5pt, black, -] (-29+48,5)--(-22.5+48,5);
\draw[line width=1.5pt, black, -] (-20.5+48,5)--(-15.5+48,5);
\draw[line width=1.5pt, black, -] (-13.5+48,5)--(-5+48,5);

\draw[line width=1.5pt, black, -] (-11+48,5) to[bend left] (-14+48,2);
\draw[line width=1.5pt, black, -] (-14+48,2)--(-24+48,2);
\draw[line width=1.5pt, black, -] (-27+48,5) to[bend right] (-24+48,2);

\end{tikzpicture}
\end{center}
\captionshift
\caption{A origin block of type $O^0$ (left) and an final block of type $F^0$ (right).}\label{figure: Origin Block 0}
\end{figure}
}
\full{\figureOriginBlockO}

\begin{OBGO}\label{definition: Origin Block 0}
An \textit{origin block of level $0$} (see \Cref{figure: Origin Block 0}), also called \textit{block of type $O^0$}, is constructed similarly to an origin block from \Cref{section: Basic Construction}. 
However, it has two additional vertices in the path of vertices that are connected with edges of weight $\eX{-1}=1$. 
Furthermore, it does not necessarily contain the origin. 
If it does not contain the origin, it instead has a start vertex~$v_{\text{start}}$.
\end{OBGO}

\begin{FBGO}\label{definition: Final Block 0}
A \textit{final block of level $0$} (see \Cref{figure: Origin Block 0}), also called \textit{block of type~$F^0$}, is a block very similar to a block of type $B^0$.
However, in contrast to a block of type $B^0$, we call the head vertex \emph{exit vertex} $v_e$ and the head side \emph{exit side}.
Additionally, the \emph{exit vertex} has no skip edge and the backbone edge incident to the exit vertex has weight $0$ and is called \emph{exit edge}.
\end{FBGO}

%

\begin{Obs}\label{lemma: Compatibility Level 0}
We can always connect the head side of a block of level $0$ to the tail side of another block of level $0$ by connecting a backbone edge to a tail vertex and a skip edge to a start vertex. This can be done independently of the types of the two blocks.
\end{Obs}

\begin{restatable}{Lem}{LemmaCostEasyOptO}\label{lemma: Costs Easy Opt 0}
The shortest path from one side of a block of level $0$ to the other side of the block that visits all vertices of the block has cost $x+2$.
\end{restatable}

\subsection{Recursive Construction}

As mentioned previously, blocks of level $i$ contain blocks of level $i-1$.
Thus, the chains of blocks of level $0$ introduced in the previous subsection form a block of level $1$ (see \Cref{figure: Three Levels} with $i=0$).

\begin{Def}[Adjacency and Incidence]
A vertex is \emph{adjacent} to a block if it is adjacent to a vertex inside of the block.
Two blocks are adjacent if a vertex contained in one block is adjacent to a vertex in the other.
A block is \emph{incident} to an edge if it contains a vertex that is incident to the edge.
\end{Def}

\begin{Def}[Traversal]\label{definition: Traversal}
\emph{Traversing} a block $B$ means walking from the block on one side of $B$ to the block on the other side of $B$, while using at least one vertex of $B$.
\end{Def}

\newcommand{\figureChainLevelIEasierNotation}{
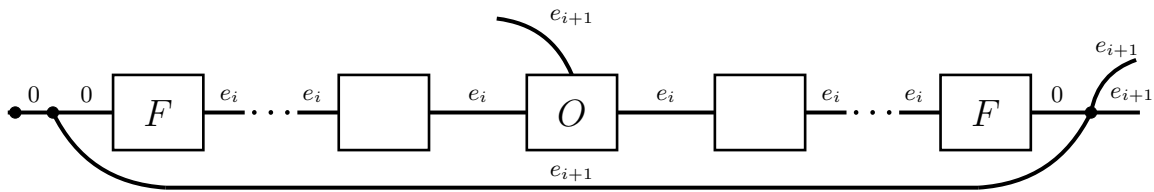
\begin{figure}\conf{[p]}
\begin{center}
\begin{tikzpicture}[xscale=.1, yscale=.1, font=\footnotesize]
\draw[line width=1pt,black,-] (-61,0) rectangle (-49,10);
\draw[line width=1pt,black,-] (-31,0) rectangle (-19,10);
\draw[line width=1pt,black,-] (-6,0) rectangle (6,10);
\draw[line width=1pt,black,-] (19,0) rectangle (31,10);
\draw[line width=1pt,black,-] (49,0) rectangle (61,10);

\draw (-71.5,5) node[above] {$0$};
\draw (-64.5,5) node[above] {$0$};
\draw (-45.5,5) node[above] {$\eX{i}$};
\draw (-34.5,5) node[above] {$\eX{i}$};
\draw (45.5,5) node[above] {$\eX{i}$};
\draw (34.5,5) node[above] {$\eX{i}$};
\draw (64.5,5) node[above] {$0$};
\draw (74.5,5) node[above] {$\eX{i+1}$};
\draw (72.5,11) node[above] {$\eX{i+1}$};
\draw (0,-5.5) node[above] {$\eX{i+1}$};

\draw (-55,5) node[font=\Large] {$F$};
\draw (0,5) node[font=\Large] {$O$};
\draw (55,5) node[font=\Large] {$F$};

\filldraw[black] (-74,5) circle (0.75);
\filldraw[black] (-69,5) circle (0.75);
\filldraw[black] (69,5) circle (0.75);

\draw (40,5) node {\Huge ...};
\draw (-40,5) node {\Huge ...};

\draw[line width=1.5pt, black, -] (61,5)--(75.5,5);
\draw[line width=1.5pt, black, -] (31,5)--(36.5,5);
\draw[line width=1.5pt, black, -] (43.5,5)--(49,5);
\draw[line width=1.5pt, black, -] (6,5)--(19,5);
\draw[line width=1.5pt, black, -] (-6,5)--(-19,5);
\draw[line width=1.5pt, black, -] (-31,5)--(-36.5,5);
\draw[line width=1.5pt, black, -] (-43.5,5)--(-49,5);
\draw[line width=1.5pt, black, -] (-61,5)--(-75,5);

\draw[line width=1.5pt, black, -] (69,5) to[bend left] (54,-5);
\draw[line width=1.5pt, black, -] (-54,-5) to[bend left] (-69,5);
\draw[line width=1.5pt, black, -] (-54,-5) -- (54,-5);
\draw[line width=1.5pt, black, -] (69,5) to[bend left] (75,12);

\draw (0,20) node[below] {$\eX{i+1}$};

\draw (-12.5,5) node[above] {$\eX{i}$};
\draw (12.5,5) node[above] {$\eX{i}$};

\draw[line width=1.5pt, black, -] (0,10) to[bend right] (-10,17.5);
\end{tikzpicture}
\end{center}
\captionshift
\caption{A series of blocks of level $i$ that form a block of level $i+1$ in simplified notation.}\label{figure: Chain Level i Easier Notation}
\end{figure}
}
\full{\figureChainLevelIEasierNotation}

We introduce blocks of level $i\in\{1,\dots,N-1\}$ inductively, where $N$ is the highest level\shorten{ (refer to \Cref{figure: Three Levels} for an example of a possible chain of level $i$)}.
Blocks of level $0<i<N$ are constructed in the same way as in level $0$ with the only difference being that vertices are replaced by subblocks of level $i-1$.
In the case $i=1$ the subblocks of level $0$ are connected according to \Cref{lemma: Compatibility Level 0}.

For convenience, we omit the skip edges in figures and label blocks with their type (see \Cref{figure: Chain Level i Easier Notation}).
Unlabeled blocks are normal blocks. We will sometimes omit the indices of blocks that denote their level if it is clear from the context.
Note that we already omitted the skip edges of the blocks of level $i-1$ in \Cref{figure: Three Levels}. 

\newcommand{\figureNormalBlockI}{
\begin{figure}\conf{[p]}
\begin{center}
\begin{tikzpicture}[scale=0.25,font=\scriptsize, decoration=brace]
\draw[line width=1pt,black,-] (-30,0)--(-10,0)--(-10,10)--(-30,10)--(-30,0);

\draw (-10,8.5) node[right] {skip,};
\draw (-10,7.5) node[right] {$\eX{i}$};
\draw (-10,2.5) node[right] {$\eX{i}$,};
\draw (-10,1.5) node[right] {backbone};
\draw[line width=1.5pt, black, -] (-11,5) -- (-5,8);
\draw[line width=1.5pt, black, -] (-11,5) -- (-5,2);

\draw (-28.5,5.5) node[above] {$0$};
\draw (-27.5,5.5) node[above] {$0$};
\draw (-29,4.5) node[below] {$v_\text{t}$};
\draw (-25.5,5.5) node[above] {$\eX{i-1}$};
\draw (-20.7,5.5) node[above] {$\eX{i-1}$};
\draw (-26,4.5) node[below] {$F_\text{t}$};
\draw (-19,4.5) node[below] {$O_\text{start}$};
\draw (-13,4.5) node[below] {$F_\text{h}$};
\draw (-18.3,5.5) node[above] {$\eX{i-1}$};
\draw (-13.5,5.5) node[above] {$\eX{i-1}$};
\draw (-11.5,5.5) node[above] {$0$};
\draw (-10.7,4.5) node[below] {$v_\text{h}$};

\draw (-19,2) node[below] {return, $\eX{i}$};

\draw (-23,5) node {\LARGE ..};
\draw (-16,5) node {\LARGE ..};

\draw[line width=1.5pt, black, -] (-29,5)--(-24.5,5);
\draw[line width=1.5pt, black, -] (-21.5,5)--(-17.5,5);
\draw[line width=1.5pt, black, -] (-14.5,5)--(-11,5);

\draw[line width=1.5pt, black, -] (-11,5) to[bend left] (-14,2);
\draw[line width=1.5pt, black, -] (-14,2)--(-25,2);
\draw[line width=1.5pt, black, -] (-28,5) to[bend right] (-25,2);

\filldraw[black] (-29,5) circle (0.25);
\filldraw[black] (-28,5) circle (0.25);
\filldraw[black] (-11,5) circle (0.25);

\foreach \x in {-27.25,-25.25,-22.25,-20.25,-18.25,-15.25,-13.25}{
	\draw[fill=white] (\x,4.25) rectangle ({\x+1.5},5.75);
}

\draw (-26.5,5) node {$F$};
\draw (-19.5,5) node {$O$};
\draw (-12.5,5) node {$F$};

\draw (-20,0) node[below] {$B^i$};

\draw (-29,0) node[below] {tail};
\draw (-11,0) node[below] {head};
\draw[decorate, yshift=2ex, thick]  (-18.25,6.75) -- node[above=0.4ex] {$\ge y$}  (-13.75,6.75);
\end{tikzpicture}
\end{center} 
\captionshift
\caption{An illustration of a normal block of type $B^i$ using the simplified notation of \Cref{figure: Chain Level i Easier Notation}.} \label{figure: Normal Block i}
\end{figure}
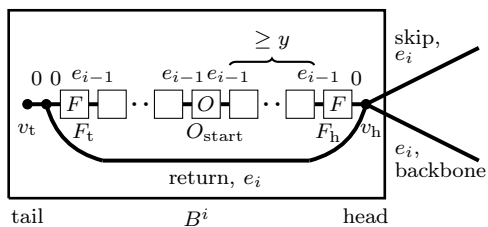
}
\full{\figureNormalBlockI}

\begin{NBGI}\label{definition: Normal Block i}
A normal block of level $i$ (see \Cref{figure: Normal Block i}), also called \textit{block of type~$B^i$}, is defined as follows.
Assume \alg starts in some subblock $O_{\text{start}}$ that is an origin block of type $O^{i-1}$ (defined below).
Once \alg leaves $O_{\text{start}}$ on one side, we let it enter a subblock that is a normal block of type $B^{i-1}$ via a skip edge.
If \alg chooses to backtrack and to continue on the other side of $O_{\text{start}}$, we add normal blocks of type $B^{i-1}$ to the other side of $O_{\text{start}}$ in the same way.
Every time \alg leaves a subblock of type~$B^{i-1}$ and takes one of the two indistinguishable edges of weight $\eX{i-1}$ at the head vertex of the subblock (after it already has examined the end point of the return edge), we let it be the skip edge, which leads to the next unvisited subblock of type~$B^{i-1}$.  
Every new subblock examined by \alg is chosen to be a normal block of type~$B^{i-1}$ until \alg has examined $x+1$ subblocks (including $O_{\text{start}}$).
Then, the unexplored subblock adjacent to the last examined subblock is chosen to be a final block of type~$F^{i-1}$ (defined below).
Likewise, the first unexplored subblock on the other end of the path of subblocks is also chosen to be a final block of type~$F^{i-1}$.
In the case that~$y$ or more subblocks lie on the path from the last examined subblock to $O_{\text{start}}$ (excluding $O_{\text{start}}$), the final block adjacent to the last examined subblock is called \emph{head subblock}~$F_\text{h}$ and the final block on the other end of the path of subblocks is called \emph{tail subblock}~$F_\text{t}$.
Otherwise, the unexplored subblock adjacent to the last examined block is called tail subblock, while the first unexplored subblock on the other end of the path of subblocks is called head subblock.
Note that in both cases there are at least~$y$ normal blocks of type~$B^{i-1}$ on the path from the~$F_\text{h}$ to $O_\text{start}$ (excluding~$F_\text{h}$ and $O_\text{start}$).
The head subblock is connected to a vertex via its backbone edge of weight $0$.
This vertex is called \emph{head vertex}~$v_h$ and it is incident to three edges of weight~$\eX{i}$.
If \alg is at the head vertex and chooses to take an edge of weight~$\eX{i}$ (which we then call \emph{return edge}), it reaches a vertex surrounded by two edges of weight $0$.
The other two edges, we call \emph{skip edge} and \emph{backbone edge}, depending on the behavior of \alg later on.
One of the two edges of weight zero is the backbone edge of the tail subblock~$F_\text{t}$.
The vertex incident to the other edge of weight~$0$ is called \emph{tail vertex}~$v_\text{t}$.
\end{NBGI}

\newcommand{\figureFinalBlockI}{
\begin{figure}\conf{[p]}
\begin{center}
\begin{tikzpicture}[scale=0.28,font=\scriptsize]
\draw[line width=1pt,black,-] (2.5-32.5,0)--(22.5-32.5,0)--(22.5-32.5,10)--(2.5-32.5,10)--(2.5-32.5,0);

\draw (22.5-32.5,8.5) node[right] {skip,};
\draw (22.5-32.5,7.5) node[right] {$\eX{i}$};
\draw (22.5-32.5,2.5) node[right] {$\eX{i}$,};
\draw (22.5-32.5,1.5) node[right] {backbone};
\draw[line width=1.5pt, black, -] (21.5-32.5,5) -- (27.5-32.5,8);
\draw[line width=1.5pt, black, -] (21.5-32.5,5) -- (27.5-32.5,2);

\draw (2.5-32.5,8.5) node[left] {skip,};
\draw (2.5-32.5,7.5) node[left] {$\eX{i}$};
\draw (2.5-32.5,2.5) node[left] {$\eX{i}$,};
\draw (2.5-32.5,1.5) node[left] {backbone};
\draw[line width=1.5pt, black, -] (3.5-32.5,5) -- (-2.5-32.5,8);
\draw[line width=1.5pt, black, -] (3.5-32.5,5) -- (-2.5-32.5,2);

\draw (3.4-32.5,5) node[below] {$v_\text{h}$};
\draw (5-32.5,4.5) node[below] {$F_\text{h}$};
\draw (4-32.5,5.5) node[above] {$0$};
\draw (6-32.5,5.5) node[above] {$\eX{i-1}$};
\draw (11.3-32.5,5.5) node[above] {$\eX{i-1}$};
\draw (12.5-32.5,4.5) node[below] {$O_\text{start}$};
\draw (13.7-32.5,5.5) node[above] {$\eX{i-1}$};
\draw (19-32.5,5.5) node[above] {$\eX{i-1}$};
\draw (21-32.5,5.5) node[above] {$0$};
\draw (20-32.5,4.5) node[below] {$F_\text{h}$};
\draw (21.8-32.5,5) node[below] {$v_\text{h}$};

\draw (8.75-32.5,5) node {\LARGE ...};
\draw (16.25-32.5,5) node {\LARGE ...};

\draw[line width=1.5pt, black, -] (3.5-32.5,5)--(6.5-32.5,5);
\draw[line width=1.5pt, black, -] (10-32.5,5)--(14-32.5,5);
\draw[line width=1.5pt, black, -] (17.5-32.5,5)--(21.5-32.5,5);


\foreach \x in {4.25,6.25,9.75,11.75,13.75,17.25,19.25}{
	\draw[fill=white] (\x-32.5,4.25) rectangle ({\x+1.5-32.5},5.75);
}

\filldraw[black] (3.5-32.5,5) circle (0.25);
\filldraw[black] (21.5-32.5,5) circle (0.25);

\draw (5-32.5,5) node {$F$};
\draw (12.5-32.5,5) node {$O$};
\draw (20-32.5,5) node {$F$};

\draw (12.5-32.5,0) node[below] {$O^i$};

\draw (3.5-32.5,0) node[below] {head};
\draw (21.5-32.5,0) node[below] {head};

\draw[line width=1pt,black,-] (-30+27.5,0)--(-10+27.5,0)--(-10+27.5,10)--(-30+27.5,10)--(-30+27.5,0);
\draw[line width=1.5pt, black, -] (-11+27.5,5) -- (-5+27.5,5);

\draw (-28.5+27.5,5.5) node[above] {$0$};
\draw (-27.5+27.5,5.5) node[above] {$0$};
\draw (-29+27.5,4.5) node[below] {$v_\text{t}$};
\draw (-25.5+27.5,5.5) node[above] {$\eX{i-1}$};
\draw (-20.7+27.5,5.5) node[above] {$\eX{i-1}$};
\draw (-26.25+27.5,4.5) node[below] {$F_\text{t}$};
\draw (-19+27.5,4.5) node[below] {$O_\text{start}$};
\draw (-12.75+27.5,4.5) node[below] {$F_\text{h}$};
\draw (-18.3+27.5,5.5) node[above] {$\eX{i-1}$};
\draw (-13.5+27.5,5.5) node[above] {$\eX{i-1}$};
\draw (-11.5+27.5,5.5) node[above] {$0$};
\draw (-10.7+27.5,4.5) node[below] {$v_\text{e}$};

\draw (-19+27.5,2) node[below] {return, $\eX{i}$};

\draw (-23+27.5,5) node {\LARGE ..};
\draw (-16+27.5,5) node {\LARGE ..};

\draw[line width=1.5pt, black, -] (-29+27.5,5)--(-24.5+27.5,5);
\draw[line width=1.5pt, black, -] (-21.5+27.5,5)--(-17.5+27.5,5);
\draw[line width=1.5pt, black, -] (-14.5+27.5,5)--(-11+27.5,5);

\draw[line width=1.5pt, black, -] (-11+27.5,5) to[bend left] (-14+27.5,2);
\draw[line width=1.5pt, black, -] (-14+27.5,2)--(-25+27.5,2);
\draw[line width=1.5pt, black, -] (-28+27.5,5) to[bend right] (-25+27.5,2);

\filldraw[black] (-29+27.5,5) circle (0.25);
\filldraw[black] (-28+27.5,5) circle (0.25);
\filldraw[black] (-11+27.5,5) circle (0.25);

\foreach \x in {-27.25,-25.25,-22.25,-20.25,-18.25,-15.25,-13.25}{
	\draw[fill=white] (\x+27.5,4.25) rectangle ({\x+1.5+27.5},5.75);
}

\draw (-26.5+27.5,5) node {$F$};
\draw (-19.5+27.5,5) node {$O$};
\draw (-12.5+27.5,5) node {$F$};

\draw (-20+27.5,0) node[below] {$F^i$};

\draw (-29+27.5,0) node[below] {tail};
\draw (-11+27.5,0) node[below] {exit};
\draw (-7.5+27.5,4.75) node[below] {exit};
\draw (-7.5+27.5,5) node[above] {$0$};
\end{tikzpicture}
\end{center}
\captionshift
\caption{An illustration of an origin block of type $O^i$ (left) and a final block of type $F^i$ (right).} \label{figure: Final Block i}
\end{figure}
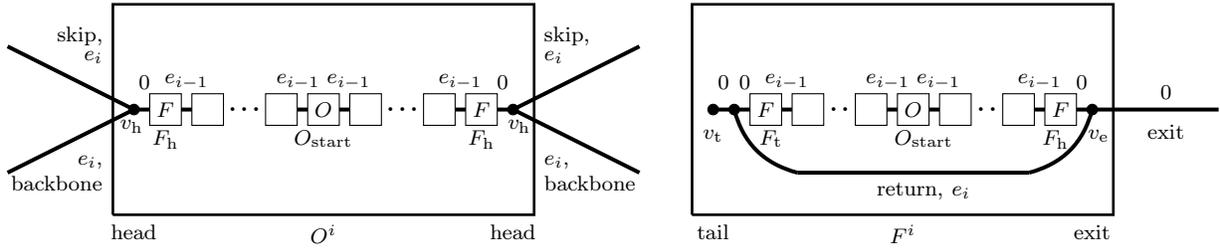
}
\full{\figureFinalBlockI}

\begin{OBGI}\label{definition: Origin Block i}
An \textit{origin block of level $i$} (see \Cref{figure: Final Block i}), also called \textit{block of type $O^i$}, is a path of $x+3$ subblocks of level $i-1$ and two vertices.
The two subblocks on the ends of the path of blocks are final blocks of type $F^{i-1}$ and are called \emph{head subblocks} $F_\text{h}$. 
Both head subblocks are connected to a vertex via their backbone edge of weight $0$.
Those two vertices are called \emph{head vertices} $v_\text{h}$. 
All edges connecting blocks with each other have weight $e_{i-1}$.
All other subblocks contained in the path are of type $B^{i-1}$ except one origin block of type $O^{i-1}$, which we call \emph{start subblock} $O_\text{start}$.
The position of $O_\text{start}$ in the path of blocks of type $B^{i-1}$ can be chosen arbitrarily.
\end{OBGI}

\begin{FBGI}\label{definition: Final Block i}
A \textit{final block of level $i$} (see \Cref{figure: Final Block i}), also called \textit{block of type $F^i$}, is a block very similar to a block of type $B^i$, i.e. is contains of $x+3$ blocks. 
However, in contrast to a block of type $B^i$, we call the head vertex \emph{exit vertex} $v_e$ and the head side \emph{exit side}.
Additionally, the \emph{exit vertex} has no skip edge and the backbone edge incident to the exit vertex has weight $0$ and is called \emph{exit edge}.
\end{FBGI}

%

\begin{restatable}{Obs}{LemmaCompatibilityRecursive}\label{lemma: Compatibility Recursive}
We can always connect the head side of a block of level $0<i<N$ to the tail side of another block of level $0<i<N$ by connecting a backbone edge to a tail vertex and a skip edge to a start subblock. This can be done independently of the types of the two blocks.
\end{restatable}
\newcommand{\LemmaCompatibilityRecursiveProof}{
\begin{proof}
We notice that every block of level $i<N$ has the same number of head sides as backbone and skip edges. Furthermore, every block has the same number of tail sides as start subblocks and tail vertices (except origin blocks, which have no tail side, but a start subblock). We can connect a head side to a tail side by connecting the backbone edge to a tail vertex and the skip edge to a start subblock. Note that every start subblock is an origin block and thus either has a start vertex or a start subblock to which we can connect the skip edge.
\end{proof}
}
\full{\LemmaCompatibilityRecursiveProof}

\begin{restatable}{Lem}{LemmaCostEasyOptI}\label{lemma: Costs Easy Opt I}
Let $B$ be any block of level $i<N$. 
Assume $B$ has backbone/exit edges incident to all tail and head vertices (i.e., by construction the block contains exactly two vertices with incident backbone/exit edges; one on each side) and assume $B$ is entered via a backbone edge or exit edge. 
Exploring the whole block and reaching the vertex adjacent to the backbone/exit edge on the other side incurs costs of exploring $(x+3)$ subblocks and walking $(x+2)$ edges of weight $\eX{i-1}$ inside $B$.
\end{restatable}

We introduce the following notation (cf. \Cref{definition: Exploration} and \Cref{definition: Traversal}):
\begin{align*}
\utX{i}&\coloneqq\text{ \alg's cost of exploring an unvisited block of type $B^i$ and entering a new unvisited block}\\ 
\tX{i}&\coloneqq\text{ cost of traversing an explored block of type $B^i$ optimally}
\end{align*}
For convenience we set $\utX{-1}:=1$, $\tX{-1}:=0$ since blocks of level $-1$ are vertices that incur no cost for exploring/traversing, but in case of $\utX{-1}$ incur a costs of $\eX{-1}=1$ for walking to the next unvisited vertex.
Furthermore, we assume that \alg can explore/traverse blocks that are not of type $B^i$ for costs of $0$.

Note that the cheapest way to traverse an explored block of type $B^i$ is to enter and leave via skip edges (recall that one skip edge is connected to the start subblock and one to the head vertex).
This way the subblocks between the tail vertex and the start subblock cam be skipped entirely.
Thus, only the subblocks between the origin subblock and the head vertex need to be traversed for a traversal.
By construction, the number of normal subblocks between the origin subblock and the head vertex is at least $y$.
Therefore, the cost of traversing an explored block of type $B^i$ is bounded by $y(\tX{i-1}+\eX{i-1})$, i.e.,
\begin{equation}\label{equation: Traversal Cost}
\tX{i}= y(\tX{i-1}+\eX{i-1})
\end{equation}
with $\tX{-1}:=0$.
Furthermore, we define the edge weights 
\begin{equation}\label{equation: Edge Cost}
\eX{i}:=x(\tX{i-1}+\eX{i-1})
\end{equation}
with $\eX{-1}:=1$.
Note that the weight $\eX{i}$ is smaller than or equal to the cost of walking from head side to the tail side of an already explored block $B^i$.

\begin{restatable}{Lem}{LemmaCostEasyLevelI}\label{lemma: Costs Easy Level I}
Let $i>0$ and $B$ be a normal block of type $B^i$ that \alg entered via skip edge, i.e., the vertex of $B$ that is first visited by \alg is inside of the start subblock.
\begin{enumerate}[topsep=0.5ex,itemsep=0.5ex,partopsep=0.5ex,parsep=0.5ex]
\item If the $(x+2)$-nd examined subblock in $B$ is the head subblock, then
\begin{itemize}[topsep=0.5ex,itemsep=0ex,partopsep=0.5ex,parsep=0.5ex]
\item \alg's cost inside $B$ until $B$ is fully explored is at least $x\utX{i-1}+\eX{i}$ and
\item \alg's last explored vertex is either one of the last two vertices on the tail side or inside of the tail subblock.
\end{itemize}
\item If the $(x+2)$-nd examined subblock in $B$ is the tail subblock, then either
\begin{itemize}[topsep=0.5ex,itemsep=0ex,partopsep=0.5ex,parsep=0.5ex]
\item \alg's cost inside $B$ until $B$ is fully explored is at least $x\utX{i-1}+(2-\frac{y}{x})\eX{i}$ and
\item \alg's last explored vertex of $B$ is the head vertex or inside of the head subblock
\end{itemize}
\vspace*{-1ex}
or
\vspace*{-1ex}
\begin{itemize}[topsep=0.5ex,itemsep=0ex,partopsep=0.5ex,parsep=0.5ex]
\item \alg's cost inside $B$ until $B$ is fully explored is at least $x\utX{i-1}+(3-\frac{y}{x})\eX{i}$ and
\item \alg's last explored vertex is either one of the last two vertices on the tail side or inside of the tail subblock.
\end{itemize}
\end{enumerate}
\end{restatable}
\newcommand{\LemmaCostEasyLevelIProof}{
\begin{proof}
%
Assume the $(x+2)$-nd subblock examined by \alg is the head subblock $F_\text{h}$.
Then there are at least $y$ subblocks between $F_\text{h}$ and $O_{\text{start}}$.
The cheapest way for \alg to be in this scenario is by walking straight in one direction and never backtrack. 
This incurs costs of at least $x\utX{i-1}$ since \alg has examined $O_{\text{start}}$ and $F_\text{h}$, has explored $x$ normal subblocks and has entered at least $x$ new unvisited blocks via skip edges.
It remains to explore the last two vertices on the tail side, the tail subblock $F_\text{t}$, the head vertex and possibly vertices in $F_\text{h}$. 
The head subblock is adjacent to the head vertex which has three incident edges of weight $\eX{i}$.
If \alg explores the head vertex, it can either backtrack to the tail subblock or take one of the edges of weight $e_i$, which by construction will be the return edge. 
Both cases add a cost of at least $\eX{i}$ (note that backtracking includes at least the traversal of $x$ explored normal blocks of type $B^{i-1}$ and $x$ edges of weight $\eX{i-1}$, i.e., has at least cost $\eX{i}$ (cf. definition (\ref{equation: Edge Cost}))).
This raises the total cost to at least $x\utX{i-1}+\eX{i}$.

Now assume the $(x+2)$-nd subblock examined by \alg is the tail subblock $F_\text{t}$.
Then there are at most $y-1$ normal subblocks between $F_\text{t}$ and $O_{\text{start}}$.
In other words, \alg has explored at least $x-y+1$ normal subblocks on the other side of $O_{\text{start}}$.
Thus, \alg has incurred costs of at least $x\utX{i-1}+(x-y+1)(\tX{i-1}+\eX{i-1})$ until until it has examined $x+2$ subblocks.
It remains to explore the last two vertices on the tail side, the head subblock and the head vertex.

If \alg explored the last two vertices on the tail side first, it incurs costs of $0$ since they are connected via $0$-weighted edges.
\alg can either backtrack to the head subblock or take the return edge of weight $\eX{i}$. 
Both cases add a cost of at least $\eX{i}$.
This raises the total cost to at least
\begin{align*}
x\utX{i-1}+(x-y+1)(\tX{i-1}+\eX{i-1})+\eX{i}&> x\utX{i-1}+(x-y)(\tX{i-1}+\eX{i-1})+\eX{i}\\
&= x\utX{i-1}+\left(2-\frac{y}{x}\right)\eX{i}
\end{align*}
and the last explored vertex is the head vertex or inside of the head subblock.

If \alg explores the head vertex and head subblock before exploring both vertices on the tail side and the tail subblock, it has to either backtrack to the head side of the block or take the return edge of weight $\eX{i}$ to explore the head vertex and head subblock.
As before, both cases add a cost of at least $\eX{i}$.
Afterwards it has to backtrack yet again to the tail side to explore the remaining vertices there.
This adds again a cost of at least $\eX{i}$, raising the total cost to at least
\begin{align*}
x\utX{i-1}+(x-y+1)(\tX{i-1}+\eX{i-1})+\eX{i}&> x\utX{i-1}+(x-y)(\tX{i-1}+\eX{i-1})+2\eX{i}\\
&= x\utX{i-1}+\left(3-\frac{y}{x}\right)\eX{i}.
\end{align*}
\end{proof}
}
\full{\LemmaCostEasyLevelIProof}

\subsection{The Highest Recursive Level}

It remains to examine the blocks of the highest level. 
The highest level is different from all other levels, because the blocks form a cycle like in \Cref{section: Basic Construction}.

\newcommand{\figureNormalBlockN}{
\begin{figure}\conf{[p]}
\begin{center}
\begin{tikzpicture}[scale=0.25,font=\scriptsize, decoration=brace]
\draw[line width=1pt,black,-] (-30,0)--(-10,0)--(-10,10)--(-30,10)--(-30,0);

\draw (-10,8.5) node[right] {skip,};
\draw (-10,7.5) node[right] {$\eX{N}$};
\draw (-10,2.5) node[right] {$\eX{N}$,};
\draw (-10,1.5) node[right] {backbone};
\draw[line width=1.5pt, black, -] (-11,5) -- (-5,8);
\draw[line width=1.5pt, black, -] (-11,5) -- (-5,2);

\draw (-28.5,5.5) node[above] {$0$};
\draw (-27.5,5.5) node[above] {$0$};
\draw (-29,4.5) node[below] {$v_\text{t}$};
\draw (-25.5,5.5) node[above] {\tiny $\eX{N-1}$};
\draw (-21,5.5) node[above] {\tiny $\eX{N-1}$};
\draw (-26,4.5) node[below] {$F_\text{t}$};
\draw (-19,4.5) node[below] {$O_\text{start}$};
\draw (-13,4.5) node[below] {$F_\text{h}$};
\draw (-18,5.5) node[above] {\tiny $\eX{N-1}$};
\draw (-13.5,5.5) node[above] {\tiny $\eX{N-1}$};
\draw (-11.5,5.5) node[above] {$0$};
\draw (-10.7,4.5) node[below] {$v_\text{h}$};

\draw (-19,2) node[below] {return, $\eX{N}$};

\draw (-23,5) node {\LARGE ..};
\draw (-16,5) node {\LARGE ..};

\draw[line width=1.5pt, black, -] (-29,5)--(-24.5,5);
\draw[line width=1.5pt, black, -] (-21.5,5)--(-17.5,5);
\draw[line width=1.5pt, black, -] (-14.5,5)--(-11,5);

\draw[line width=1.5pt, black, -] (-11,5) to[bend left] (-14,2);
\draw[line width=1.5pt, black, -] (-14,2)--(-25,2);
\draw[line width=1.5pt, black, -] (-28,5) to[bend right] (-25,2);

\filldraw[black] (-29,5) circle (0.25);
\filldraw[black] (-28,5) circle (0.25);
\filldraw[black] (-11,5) circle (0.25);

\foreach \x in {-27.25,-25.25,-22.25,-20.25,-18.25,-15.25,-13.25}{
	\draw[fill=white] (\x,4.25) rectangle ({\x+1.5},5.75);
}

\draw (-26.5,5) node {$F$};
\draw (-19.5,5) node {$O$};
\draw (-12.5,5) node {$F$};

\draw (-20,0) node[below] {$B^N$};

\draw (-29,0) node[below] {tail};
\draw (-11,0) node[below] {head};
\end{tikzpicture}
\end{center} 
\captionshift
\caption{An illustration of a normal block of type $B^N$ using the simplified notation of \Cref{figure: Chain Level i Easier Notation}.} \label{figure: Normal Block N}
\end{figure}
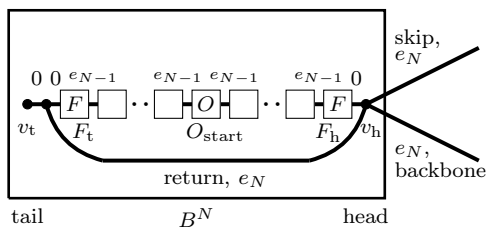
}
\full{\figureNormalBlockN}

\begin{NBGN}\label{definition: Normal Block n}
A \emph{normal block of level $N$} (see \Cref{figure: Normal Block N}), also called \emph{block of type $B^N$} is defined as follows.
Assume \alg starts in some origin block $O_{\text{start}}$ of type $O^{N-1}$.
Once \alg leaves $O_{\text{start}}$ on one side, we let it enter a subblock of type $B^{N-1}$ via a skip edge.
If \alg chooses to backtrack and to continue on the other side of $O_{\text{start}}$, we add normal blocks to the other side of $O_{\text{start}}$ in the same way.
Every time \alg leaves a subblock of type $B^{N-1}$ and takes one of the two indistinguishable edges of weight $e_{N-1}$ at the head vertex (after it already has examined the end point of the return edge), it is chosen to be the skip edge which leads to the next unvisited subblock of type $B^{N-1}$. 
Every new block examined by \alg is chosen to be a normal block of type $B^{N-1}$ until \alg has examined $x+1$ blocks (including the origin block).
The $(x+2)$-nd explored block is chosen to be a final block of type $F^{N-1}$ and is called \emph{head subblock} $F_\text{h}$ and the first unexplored block on the other end of the path of blocks is chosen to be a final subblock of type $F^{N-1}$ as well and is called the \emph{tail subblock} $F_\text{t}$.
The head subblock is connected to a vertex via its backbone edge of weight $0$.
This vertex is called the \emph{head vertex} $v_h$ and it is incident to three edges of weight~$e_N$.
If \alg chooses to take an edge of weight $e_{N}$ (which we then call \emph{return edge}), it reaches a vertex surrounded by two edges of weight $0$.
The other two edges, we call \emph{skip edge} and \emph{backbone edge}, depending on the behavior of \alg later on.
One of the two edges of weight zero is the backbone edge of the tail subblock $F_\text{t}$.
The vertex incident to the other edge of weight $0$ is called \emph{tail vertex}~$v_\text{t}$.
\end{NBGN}

Note that we omit the requirement of having at least $y$ blocks between the origin subblock and the head subblock inside a normal block of level $N$.

\begin{restatable}{Lem}{LemmaCostsEasyRecursive}\label{lemma: Costs Easy Recursive}
Let $B$ be a normal block of type $B^N$ and $y_N\in\{0,\dots,x\}$ be the number of normal blocks between the origin subblock and the head subblock of $B$. 
The total cost of exploring $B$ is at least $x\utX{N-1}+(x-y_N-1)(\tX{N-1}+\eX{N-1})+\eX{N}$ and the last explored vertex is either one of the last two vertices on the tail side or inside of the tail subblock.
\end{restatable}
\newcommand{\LemmaCostsEasyRecursiveProof}{
\begin{proof}
By definition, in a normal block of type $B^N$ the $(x+2)$-nd examined subblock is always the head subblock. 
Thus, the proof is the same as for \Cref{lemma: Costs Easy Level I} Case $1$ except one difference:
If there are $y_N\in\{0,\dots,x\}$ normal subblocks between the origin subblock and the head subblock of $B$, the agent has examined $x-y_N$ normal subblocks on one side of the origin subblock, then has backtracked to the origin subblock and has examined $y_N$ normal subblocks on the other side before visiting the head subblock.
This backtracking to the origin subblock adds additional costs of $(x-y_N)(\tX{N-1}+\eX{N-1})$.
\end{proof}
}
\full{\LemmaCostsEasyRecursiveProof}

\begin{OBGN}\label{definition: Origin Block n}
An \textit{origin block of the level $N$}, also called \textit{block of type $O^N$} is defined as block of type $O^i$ with $i=N$.
\end{OBGN}

\begin{FBGN}\label{definition: Final Block n}
A \textit{final block of the level $N$}, also called \textit{block of type $F^N$} is defined as block of type $F^i$ with $i=N$.
\end{FBGN}

\newcommand{\figureFinalBlockN}{
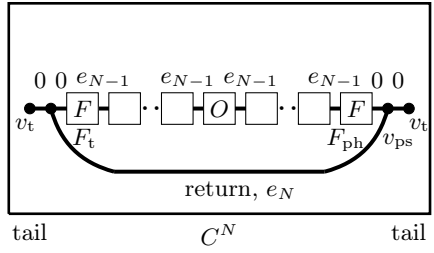
\begin{figure}\conf{[p]}
\begin{center}
\begin{tikzpicture}[scale=0.28,font=\footnotesize]
\draw[line width=1pt,black,-] (-10,0)--(10,0)--(10,10)--(-10,10)--(-10,0);

\draw (-9.2,5) node[below] {$v_\text{t}$};
\draw (-8.5,5.5) node[above] {$0$};
\draw (-7.5,5.5) node[above] {$0$};
\draw (-6.4,4.5) node[below] {$F_\text{t}$};
\draw (-5.5,5.5) node[above] {$e_{N-1}$};
\draw (-1.5,5.5) node[above] {$e_{N-1}$};
\draw (1.5,5.5) node[above] {$e_{N-1}$};
\draw (5.5,5.5) node[above] {$e_{N-1}$};
\draw (7.5,5.5) node[above] {$0$};
\draw (8.5,5.5) node[above] {$0$};
\draw (6,4.5) node[below] {$F_\text{ph}$};
\draw (8.5,4.3) node[below] {$v_\text{ps}$};
\draw (9.5,5) node[below] {$v_\text{t}$};
\draw (1,2) node[below] {return, $e_N$};

\draw[line width=1.5pt, black, -] (-9,5)--(-5,5);
\draw[line width=1.5pt, black, -] (-2.5,5)--(2.5,5);
\draw[line width=1.5pt, black, -] (5,5)--(9,5);

\draw[line width=1.5pt, black, -] (8,5) to[bend left] (5,2);
\draw[line width=1.5pt, black, -] (5,2)--(-5,2);
\draw[line width=1.5pt, black, -] (-8,5) to[bend right] (-5,2);

\filldraw[black] (-9,5) circle (0.25);
\filldraw[black] (-8,5) circle (0.25);
\filldraw[black] (8,5) circle (0.25);
\filldraw[black] (9,5) circle (0.25);

\draw (-3.25,5) node {\LARGE ..};
\draw (3.25,5) node {\LARGE ..};

\foreach \x in{-7.25,-5.25,-2.75,-0.75,1.25,3.75,5.75}{
	\draw[fill=white] (\x,4.25) rectangle ({\x+1.5},5.75);
}

\draw (-6.5,5) node {$F$};
\draw (0,5) node {$O$};
\draw (6.5,5) node {$F$};

\draw (0,0) node[below] {$C^N$};

\draw (-9,0) node[below] {tail};
\draw (9,0) node[below] {tail};
\end{tikzpicture}
\end{center}
\captionshift
\caption{An illustration of a closing block gadget of type $C^N$.} \label{figure: Final Block n}
\end{figure}
}
\full{\figureFinalBlockN}

\begin{CBGN}\label{definition: Closing Block n}
A \textit{closing block of level $N$} (See \Cref{figure: Final Block n}), also called \textit{block of type $C^N$}, is a path of $x+3$ subblocks of level $N-1$ and four vertices.
The two subblocks on the ends of the path of blocks are final blocks of type $F^{N-1}$ and one is called \emph{tail subblock} $F_\text{t}$ and the other one is called \emph{pseudo head subblock} $F_\text{ph}$. 
Both final blocks are connected to a vertex via their backbone edge of weight $0$.
The vertex connected to $F_\text{ph}$ is called \emph{pseudo start vertex} $v_\text{ps}$ and it is connected to the vertex adjacent to $F_\text{t}$ via a \emph{return edge} of weight $e_{N}$.
Furthermore, $v_\text{ps}$ is connected to another vertex via an edge of weight $0$ and the vertex adjacent to $F_\text{t}$ is connected to yet another vertex via an edge of weight $0$.
Those two vertices are both called \emph{tail vertex} $v_\text{t}$.
All edges connecting blocks with each other have weight $e_{N-1}$.
All subblocks besides $F_\text{t}$ and $F_\text{ph}$ contained in the path are of type $B^{N-1}$ except one origin block of type $O^{N-1}$, which we call \emph{start subblock} $O_\text{start}$.
The position of $O_\text{start}$ in the path of blocks of type $B^{i-1}$ is dependent on the behavior of \alg.
\end{CBGN}

\begin{restatable}{Obs}{LemmaCompatibilityRecursiveN}\label{lemma: Compatibility Recursive N}
We can always connect the head side of a block of level $N$ to the tail side of another block of level $N$ by connecting the backbone edge to the tail vertex and the skip edge to the (pseudo) start subblock/vertex. This can be done independently of the types of the two blocks.
\end{restatable}
\newcommand{\LemmaCompatibilityRecursiveNProof}{
\begin{proof}
Every block has the same number of head sides as backbone and skip edges and the same number of tail sides as start subblocks/(pseudo) start vertices and tail vertices. To complete the proof, we need to check if the claimed connections work: Since start subblocks are always origin blocks we can connect the skip edges to their contained start subblocks/start vertices.
\end{proof}
}
\full{\LemmaCompatibilityRecursiveNProof}

\subsection{Analysis of the Recursive Costs}

We describe the macroscopic design of the graph $G_{\text{rec}}$, i.e., how the blocks in the highest level $N$ are connected to each other.

\begin{GThree}\label{strategy: Simple Circle Recursive}
We construct $G_{\text{rec}}$ similarly to $G_{\text{simple}}$:
\alg starts at the origin inside an origin block of type~$O^0$, which is (through multiple levels) contained in an origin block of type~$O^N$.
We let \alg explore the block until it reaches a head vertex leading outside of the origin block.
If \alg takes one of the two edges of weight~$\eX{N}$ at a head vertex of the origin block, we fix it to be the skip edge.
Independently of the head side chosen by \alg, we let \alg enter a normal block of type~$B^N$.
If, on the other hand, \alg chooses to backtrack and continue on the other side of the origin, we add normal blocks to the other side of the origin block the same way.
Note that we can connect blocks of type~$B^N$ to both sides of the origin block, by \Cref{lemma: Compatibility Recursive N}.
If \alg leaves a normal block~$B$ of type~$B^N$ via the skip edge it used to enter~$B$ before it has visited the head vertex
and reenters~$B$ via the backbone edge, we connect one end of the explored path in B to a final subblock of type~$F^{N-1}$ that is adjacent to the vertex, which is adjacent to the tail vertex of~$B$.
Every time \alg is at a head vertex of a normal block and takes one of the two remaining (it already took the return edge; cf. construction of normal of type $B^N$) edges of weight $e_N$, we let it be the skip edge. 
Every new block examined by \alg is chosen to be a normal block of type~$B^N$ until \alg has examined $x+2$ blocks (including the origin block).
The $(x+3)$-nd block examined by \alg is chosen to be a closing block of type~$C^N$ connecting both ends of our path of blocks.
\end{GThree}

Note that $G_\text{rec}$ is a planar graph: 
As with $G_\text{simple}$, if we exclude the skip and return edges, $G_\text{rec}$ is just a cycle.
If we embed this cycle in the plane, we can add the skip edges on the outer side of the cycle and the return edges on the inner side of the cycle.
As before, the planarity of $G_\text{rec}$ then follows from the fact that skip edges and return edges can be drawn without crossing.

It remains to bound the costs for \alg and \opt in a normal block of type $B^N$ and eventually in~$G_\text{rec}$.
For this, we need to analyze recursive formulas for the costs of \alg and \opt.

Recall that for simplicity we assume that \alg is able to explore every block except normal blocks of type $B^i$ for costs of $0$.
Thus, we only examine the exploration of normal blocks in detail.

For the optimum cost, we introduce the following notation (cf. \Cref{definition: Exploration}):
\begin{equation*}
\vX{i}\coloneqq\text{ \opt's cost of exploring an unvisited block level $i$ and entering a new unvisited block}\\
\end{equation*}
For convenience we set $\vX{-1}:=1$ since blocks of level $-1$ are vertices that incur no cost for exploring, but incur a costs of $\eX{-1}=1$ for walking to the next unvisited vertex.
Note that $\vX{i}$ is well-defined since the different kinds of blocks can all be explored optimally by omitting all skip and return edges and just moving from one side to another.
Thus, by \Cref{lemma: Costs Easy Opt 0} and \Cref{lemma: Costs Easy Opt I} we get the following result.
\begin{restatable}{Lem}{LemmaCostEasyOptFinal}\label{lemma: Costs Easy Opt Final}
Let $0\le i\le N$. \opt's cost of exploring a block level $i$ and entering an adjacent block is
\begin{equation*}
\vX{i}=(x+3)\vX{i-1}-\eX{i-1}+\eX{i}.
\end{equation*}
\end{restatable}
\newcommand{\LemmaCostEasyOptFinalProof}{
\begin{proof}
\Cref{lemma: Costs Easy Opt I} also holds for all $0\le i\le N$.
Thus we have $(x+3)$ times the cost of exploring an unvisited subblock of level $i-1$ and $(x+2)$ times the cost of entering a new unvisited block.
This incurs costs of $(x+3)\vX{i-1}-\eX{i-1}$.
Since the agent still has to enter a new block via backbone edge, an additional cost of $\eX{i}$ is added.
\end{proof}
}
\full{\LemmaCostEasyOptFinalProof}

\begin{restatable}{Lem}{LemmaCostPerBlockI}\label{lemma: Lemma Cost Per Block I}
Suppose $B$ is a normal block of level $i<N$ that has been entered by \alg via the skip edge of a block next to the tail side of $B$ for the first time. 
Then \alg incurs a cost of at least 
\begin{equation*}
\utX{i}\ge x\utX{i-1}+\left(3-\frac{y}{x}\right)\eX{i}
\end{equation*}
inside $B$ until it has explored $B$ and has entered the unexplored block adjacent to the head side of $B$. 
In the case that $B$ is a normal block of level $N$, \alg incurs a cost of at least 
\begin{equation*}
\utX{N}\ge x\utX{N-1}+(x-y_N)(\tX{N-1}+\eX{N-1})+3\eX{N}
\end{equation*}
inside $B$, where $y_N\in\{0,\dots, x\}$ is the number of normal blocks between the origin subblock and the head subblock of $B$.
\end{restatable}
\newcommand{\LemmaCostPerBlockIProof}{
\begin{proof}
Let $i\in\{0,\dots,N\}$.
First, we notice that since \alg enters $B$ via a skip edge, the block adjacent to the tail side of $B$ is already examined by \Cref{definition: Examination} (but not necessarily explored).

Assume \alg leaves~$B$ before exploring it and instead goes back through the skip edge it first arrived by. 
This incurs a cost of~$\eX{i}$.
For exploring~$B$, \alg has to reenter~$B$ either by walking the skip edge or by walking the backbone edge.
Both options incur a cost of~$\eX{i}$.
Exploring every subblock of~$B$ then incurs an additional cost of even more than $x\utX{i-1}$. 
Finally, \alg has to leave~$B$ to enter a new unexplored block.
This again incurs a cost of at least~$\eX{i}$.
Thus, in total, \alg has a cost of at least $x\utX{i-1}+3\eX{i}$.
Note that $B$ cannot be entered from the head side, since then $B$ would be the $(x+2)$-nd examined block of level $N$ in $G_\text{rec}$ and by definition a closing block.

Therefore, we may assume that \alg stays inside~$B$ and explores~$B$.
Now let $i<N$.
According to \Cref{lemma: Costs Easy Level I} we have to consider two cases.
If \alg's last explored vertex is one of the two vertices on the tail side of~$B$ or inside of the tail subblock of~$B$, it has costs of at least $x\utX{i-1}+\eX{i}$ for exploring~$B$.
In the case that \alg now leaves~$B$ via the backbone edge adjacent to the tail vertex of $B$, it incurs a cost of $\eX{i}$.
Since the block adjacent to the tail side of~$B$ is already examined, \alg has to do at least one more transition to another block incurring again a cost of $\eX{i}$ and resulting in a total cost of at least $x\utX{i-1}+3\eX{i}$. 
Otherwise \alg backtracks to the head side of~$B$ for costs of at least $\eX{i}$ and enters the unexamined block adjacent to the head side of $B$ by taking either the skip or the backbone edge adjacent to the head vertex of~$B$ for a cost of $\eX{i}$, also resulting in a total cost of at least $x\utX{i-1}+3\eX{i}$.

If \alg's last explored vertex is the head vertex on the head side of $B$, it has costs of at least $x(\utX{i-1}+\eX{i-1})+(2-\frac{y}{x})\eX{i}$ for exploring $B$ according to \Cref{lemma: Costs Easy Level I}.
It remains to enter a new unexplored block by taking either the skip or the backbone edge adjacent to the head vertex of $B$ for a cost of $\eX{i}$, resulting in a total cost of at least $x\utX{i-1}+(3-\frac{y}{x})\eX{i}$.

If $i=N$, according to \Cref{lemma: Costs Easy Recursive}, \alg has costs of at least $x\utX{N-1}+(x-y_N)(\tX{N-1}+\eX{N-1})+\eX{N}$ for exploring~$B$ and its last explored vertex is one of the two vertices on the tail side of $B$ or inside of the tail subblock of~$B$.
Since the block adjacent to the tail side of~$B$ is already examined \alg has to do at least one more transition to another block incurring again a cost of~$\eX{N}$ and resulting in a total cost of at least $x\utX{N-1}+(x-y_N)(\tX{N-1}+\eX{N-1})+3\eX{N}$. 
Otherwise \alg backtracks to the head side of~$B$ for costs of at least~$\eX{N}$ and enters the unexamined block adjacent to the head side of~$B$ by taking either the skip or the backbone edge adjacent to the head vertex of $B$ for a cost of $\eX{N}$, also resulting in a total cost of at least $x\utX{N-1}+(x-y_N)(\tX{N-1}+\eX{N-1})+3\eX{N}$.

\end{proof}
}
\full{\LemmaCostPerBlockIProof}

We define 
\begin{equation}
\uX{i}\coloneqq x\uX{i-1}+\left(3-\frac{y}{x}\right)\eX{i}\label{equation: U General}
\end{equation}
for $i\ge 0$ with $\uX{-1}\coloneqq 0$.
Note that we have 
\begin{equation}
\utX{i}\ge \uX{i}\label{Urec bigger U}
\end{equation}
for all $0\le i\le N$, i.e., $\uX{i}$ is a lower bound on the costs of \alg for exploring a normal block of type $B^i$.
We will use $\utX{i}$ instead of $\uX{i}$ in the next section since it allows an easier analysis of $G_\text{rec}$.

\subsection{Analysis of the Competitive Ratio}

It remains to compute the ratio between \alg's costs exploring $G_\text{rec}$ and the optimum costs.
First we examine the degree of the polynomials $\uX{i}$, $\vX{i}$, $\tX{i}$ and $\eX{i}$ with respect to $x$.

\begin{restatable}{Lem}{LemmaDegreeOfPols}\label{lemma: Degree of Pols}
Let $i\ge 0$. We have
\begin{equation*}
\uX{i}\in\Theta(x^{i+1}),\qquad \vX{i}\in\Theta(x^{i+1}),\qquad \eX{i}\in\Theta(x^{i+1}),\qquad \tX{i}\in O(x^{i+1}).
\end{equation*}
\end{restatable}
\newcommand{\LemmaDegreeOfPolsProof}{
\begin{proof}
We prove the statement inductively. 
According to equation \ref{equation: U General} we have $\uX{0}= 4x-y$ since $\eX{0}=x$ (see equation \ref{equation: Edge Cost}).
Furthermore, we have $\vX{0}=2x+2$ (\Cref{lemma: Costs Easy Opt Final}) and $\tX{0}=y$ (equation \ref{equation: Traversal Cost}).
Since we have $y\in\{0,\dots,\lfloor\frac{x}{2}\rfloor\}$, i.e., $y\in O(x)$, the claim is true for $i=0$.
Now assume, the claim is true for $i-1$.
We have 
\begin{equation*}
\eX{i}\overset{(\ref{equation: Edge Cost})}{=}x(\eX{i-1}+\tX{i-1})\in x\cdot \Theta(x^i)+x\cdot O(x^i)=\Theta(x^{i+1})
\end{equation*}
and
\begin{equation*}
\tX{i}\overset{(\ref{equation: Traversal Cost})}{=}y(\eX{i-1}+\tX{i-1})\in y\cdot \Theta(x^i)+y\cdot O(x^i)\subseteq O(x^{i+1}).
\end{equation*}
Thus, the claim is true for $\eX{i}$ and $\tX{i}$. 
Furthermore, we have 
\begin{equation*}
\vX{i}\overset{\text{Lem. }\ref{lemma: Costs Easy Opt Final}}{=}(x+3)\vX{i-1}-\eX{i-1}+\eX{i}\in (x+3)\Theta(x^i)-\Theta(x^i)+\Theta(x^{i+1})=\Theta(x^{i+1}).
\end{equation*}
and
\begin{align*}
\uX{i}&\myoverset{(\ref{equation: U General})}{=}{15}x\uX{i-1}+(3-\tfrac{y}{x})\eX{i}\\
&\myoverset{}{\in}{15} x\cdot\Theta(x^{i})+(3-\tfrac{y}{x})\Theta(x^{i+1})\\
&\myoverset{}{=}{15} \Theta(x^{i+1}),
\end{align*}
i.e., the claim is also true for $\vX{i}$ and $\uX{i}$.
\end{proof}
}
\full{\LemmaDegreeOfPolsProof}

\begin{restatable}{Lem}{LemmaTwoTimesEdge}\label{lemma: Two Times Edge}
Let $i\ge 0$. We have $\eX{i}=(x+y)\eX{i-1}$.
\end{restatable}
\newcommand{\LemmaTwoTimesEdgeProof}{
\begin{proof}
We have 
\begin{equation*}
\eX{i} = x(\eX{i-1}+\tX{i-1})\qquad\text{ and }\qquad \tX{i-1} = y(\eX{i-2}+\tX{i-2})=\frac{y}{x}\eX{i-1}
\end{equation*}
and thus
\begin{equation*}
\eX{i} = x\left(\eX{i-1}+\frac{y}{x}\eX{i-1}\right)=(x+y)\eX{i-1}.
\end{equation*}
\end{proof}
}
\full{\LemmaTwoTimesEdgeProof}

\begin{restatable}{Lem}{LemmaRepresentationUOne}\label{lemma: Representation U One}
Let $i\ge 0$ and $B$ be a normal block of type $B^i$. We have
\begin{equation*}
\uX{i}= x^{i+1}\uX{-1}+\sum_{j=0}^ix^j(3-\tfrac{y}{x})\eX{i-j}.
\end{equation*}
\end{restatable}
\newcommand{\LemmaRepresentationUOneProof}{
\begin{proof}
We show the claim by showing inductively that for every $0\le k \le i$ we have
\begin{equation}
\uX{i} = x^{k+1}\uX{i-k-1}+\sum_{j=0}^kx^j(3-\tfrac{y}{x})\eX{i-j}.\label{equation: Induction}
\end{equation}
For $k=0$ we have
\begin{equation*}
\uX{i} \overset{(\ref{equation: U General})}{=} x\uX{i-1}+(3-\tfrac{y}{x})\eX{i}.
\end{equation*}
Now assume the claim is true for $k-1$.
Then we have
\begin{align*}
\uX{i} &\myoverset{}{=}{35} x^{k}\uX{i-(k-1)-1}+\sum_{j=0}^{k-1}x^j(3-\tfrac{y}{x})\eX{i-j}\\
&\myoverset{(\ref{equation: U General})}{=}{35} x^{k}(x\uX{i-k-1}+(3-\tfrac{y}{x})\eX{i-k})+\sum_{j=0}^{k-1}x^j(3-\tfrac{y}{x})\eX{i-j}\\
&\myoverset{}{=}{35} x^{k+1}\uX{i-k-1}+x^{k}(3-\tfrac{y}{x})\eX{i-k}+\sum_{j=0}^{k-1}x^j(3-\tfrac{y}{x})\eX{i-j}\\
&\myoverset{}{=}{35} x^{k+1}\uX{i-k-1}+\sum_{j=0}^{k}x^j(3-\tfrac{y}{x})\eX{i-j},
\end{align*}
i.e., the equality (\ref{equation: Induction}) holds for $k$. 
The claim now follows since (\ref{equation: Induction}) also holds for $k=i$.
\end{proof}
}
\full{\LemmaRepresentationUOneProof}

If we choose $y=0$ the ratio of \alg's costs of exploring $G_\text{rec}$ and the optimum cost is largest.
This is the case since \alg is only forced to do a moderate amount of backtracking in $G_\text{rec}$.
In \Cref{section: Further Improvements to the Lower Bound} we will present a graph, where the best choice is $y>0$.
For now we assume $y=0$.

\begin{restatable}{Lem}{LemmaUELargerVE}\label{lemma: UE Larger VE}
Let $i\ge 0$ and $y=0$. We have
\begin{equation*}
\uX{i}= \left(3-\frac{2}{i+2}\right)\vX{i}-O(x^i).
\end{equation*}
\end{restatable}
\newcommand{\LemmaUELargerVEProof}{
\begin{proof}
We have 
\begin{align*}
\uX{0}&\myoverset{(\ref{equation: U General})}{=}{15} 4x\\
&\myoverset{}{=}{15} 2(2x+2)-4\\
&\myoverset{}{=}{15} 2\vX{0}-O(1).
\end{align*}
Thus, the claim holds for $i=0$.
Now let $i\ge 1$. We show the claim by showing the equation
\begin{equation}
\uX{i}= x^{i-k}\left(3-\frac{2}{i+2}\right)\vX{k}+\sum_{j=0}^{i-k-1}x^j\left(3-\frac{2}{i+2}\right)\eX{i-j}-O(x^i)\label{equation: Induction Two}
\end{equation}
inductively.
Let $k=0$.
We have
\begin{align*}
\uX{i}&\myoverset{Lem. \ref{lemma: Representation U One}}{=}{90} x^{i+1}\uX{-1}+\sum_{j=0}^i3x^j\eX{i-j}\\
&\myoverset{$\uX{-1}=\vX{-1}$}{=}{90} x^{i+1}\vX{-1}+3x^{i}\eX{0}+\sum_{j=0}^{i-1}3x^j\eX{i-j}\\
&\mybothset{Lem. \ref{lemma: Costs Easy Opt Final}}{=}{$\vX{0}=x\vX{-1}+\eX{0}+O(1)$}{90}{90} x^i\vX{0}+2x^i\eX{0}+\sum_{j=0}^{i-1}3x^j\eX{i-j}-O(x^i)\\
&\mybothset{Lem. \ref{lemma: Two Times Edge}}{=}{$\eX{j}= x^j\eX{0}$}{90}{90} x^i\vX{0}+\left(2+\frac{2i}{i+2}\right)x^i\eX{0}+\sum_{j=0}^{i-1}\left(3-\frac{2}{i+2}\right)x^j\eX{i-j}-O(x^i)\\
&\myoverset{}{=}{90} x^i\vX{0}+\left(4-\frac{4}{i+2}\right)x^i\eX{0}+\sum_{j=0}^{i-1}\left(3-\frac{2}{i+2}\right)x^j\eX{i-j}-O(x^i)\\
&\myoverset{$\vX{0}=2\eX{0}+O(1)$}{=}{90} x^i\left(3-\frac{2}{i+2}\right)\vX{0}+\sum_{j=0}^{i-1}\left(3-\frac{2}{i+2}\right)x^j\eX{i-j}-O(x^i)
\end{align*}
as claimed.
Now assume equation (\ref{equation: Induction Two}) is satisfied for $k-1$.
We have 
\begin{align*}
\uX{i}&\myoverset{(\ref{equation: Induction Two})}{=}{95}x^{i-k+1} \left(3-\frac{2}{i+2}\right)\vX{k-1}+\sum_{j=0}^{i-k}\left(3-\frac{2}{i+2}\right)x^j\eX{i-j}-O(x^i)\\
&\mybothset{Lem. \ref{lemma: Costs Easy Opt Final}}{=}{$\vX{k}=x\vX{k-1}+\eX{k}+O(x^k)$}{95}{95} x^{i-k}\left(3-\frac{2}{i+2}\right)\vX{k}+\sum_{j=0}^{i-k-1}\left(3-\frac{2}{i+2}\right)x^j\eX{i-j}-O(x^i),\\
\end{align*}
i.e., equality (\ref{equation: Induction Two}) is satisfied for $k$.
The claim now follows from the fact that equation (\ref{equation: Induction Two}) is satisfied for $k=i$.
\end{proof}
}
\full{\LemmaUELargerVEProof}

\newcommand{\TheoremLowerBoundThreeProof}{
\begin{proof}[Proof of \Cref{theorem: Lower Bound 3}]
We can apply \Cref{lemma: Lemma Cost Per Block I} for at least~$x$ of the~$x+1$ normal blocks of type~$B^N$, accumulating a cost of at least $\uX{N}$ per application of \Cref{lemma: Lemma Cost Per Block I} by equation (\ref{Urec bigger U}). 
Note that \Cref{lemma: Lemma Cost Per Block I} might not be applicable for one normal block that is connected to the closing block since the closing block can be explored from two sides.
Thus, in total, we have
\begin{equation}
\alg(G_\text{rec})\ge x\uX{N}.\label{equation: Alg Rec}
\end{equation}
The optimum has to explore $x+3$ blocks and has to walk $x+3$ edges of weight $\eX{N}$.
Thus,
\begin{equation}
\opt(G_\text{rec})=(x+3)\vX{N}=x\vX{N}+O(x^{N+1}).\label{equation: Opt Rec}
\end{equation}
The competitive ratio is
\begin{align*}
f(x)&\myoverset{}{\coloneqq}{40}\frac{\alg(G_\text{rec})}{\opt(G_\text{rec})}\\
&\myoverset{(\ref{equation: Alg Rec}),(\ref{equation: Opt Rec})}{\ge}{40}\frac{x\uX{N}{}}{x\vX{N}{}+O(x^{N+1})}\\
&\myoverset{Lem. \ref{lemma: UE Larger VE}}{\ge}{40}\frac{x\left(3-\frac{2}{N+2}\right)\vX{N}-O(x^{N+1})}{x\vX{N}+O(x^{N+1})}\\
\end{align*}
Note that we have 
\begin{equation*}
x\left(3-\frac{2}{N+2}\right)\vX{N}\in\Theta(x^{N+2})\quad\text{and}\quad x\vX{N}\in O(x^{N+2}).
\end{equation*}
by \Cref{lemma: Degree of Pols}.
Thus letting the number of blocks per block $x$ as well as the number of levels $N$ go to infinity, we get
\begin{equation*}
f(x)\ge 3-\frac{2}{N+2}\overset{N\rightarrow \infty}{\longrightarrow}3>3-\varepsilon.
\end{equation*}
\end{proof}
}
\full{\TheoremLowerBoundThreeProof}

\section{Further Improvements to the Lower Bound}\label{section: Further Improvements to the Lower Bound}

Even though the construction presented in \Cref{section: Improved Lower Bound of 3} has a quite complex recursive structure, from a macroscopic point of view, the resulting graph is simply a cycle.
In this section, we replace the cycle structure by a series of cycles (see \Cref{figure: Basic Ring Structure}). 
We call the resulting graph $G_\text{chain}$.
In this graph, \alg is forced to traverse certain blocks twice, while the optimum is still able to traverse the entire graph and return to the origin traversing every block only once. 
The recursive construction of the blocks itself remains the same. This approach further increases the lower bound on the competitive ratio to $\frac{10}{3}$.
In particular, we will show the following.

\begin{restatable}{The}{TheoremLowerBoundFour}\label{theorem: Lower Bound 4}
For every $\varepsilon>0$ we can construct a graph $G_\text{chain}$ such that 
\begin{equation*}
\alg(G_\text{chain})\ge \left(\frac{10}{3}-\varepsilon\right)\cdot\opt(G_\text{chain}).
\end{equation*}
\end{restatable}

\newcommand{\figureBasicRingStructure}{
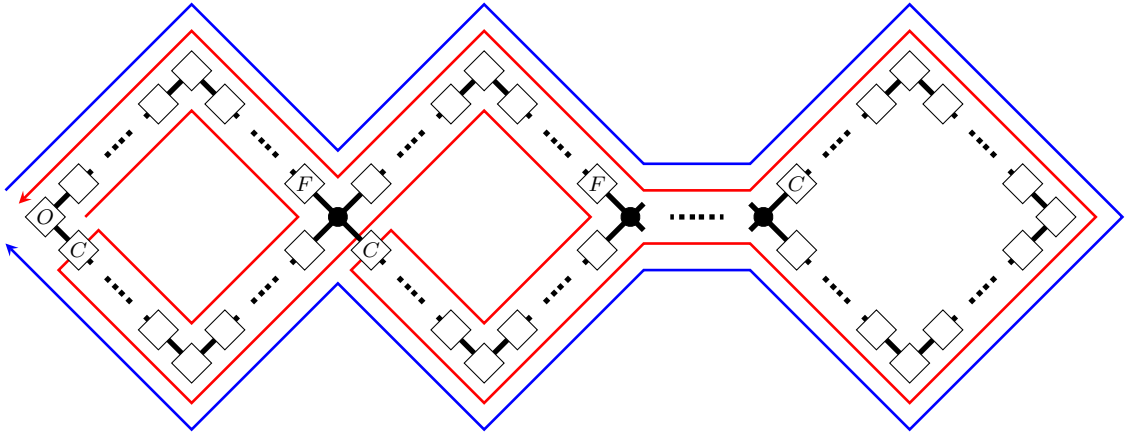
\begin{figure}\conf{[p]}
\begin{center}
\begin{tikzpicture}[scale=0.25,font=\normalsize, rotate=45]

\draw[line width=1pt,blue,->, >=stealth] (-20,13)--(-6,13)--(-6,2)--(5,2)--(5,-10)--(9,-14)--(21,-14)--(21,-30)--(5,-30)--(5,-18)
--(1,-14)--(-11,-14)--(-11,-3)--(-22,-3)--(-22,11);

\draw[line width=1pt,red,->, >=stealth] (-18,9)--(-10,9)--(-10,1)--(-18,1)--(-18,8)--(-21,8)--(-21,-2)--(1,-2)
--(1,-10)--(-7,-10)--(-7,-3)--(-10,-3)--(-10,-13)--(2,-13)--(6,-17)--(6,-29)--(20,-29)--(20,-15)--(8,-15)--(4,-11)
--(4,1)--(-7,1)--(-7,12)--(-20,12);

\draw[line width=2pt,black,dotted,-] (-13.9,10.5)--(-13,10.5);
\draw[line width=2pt,black,dotted,-] (-14.1,10.5)--(-15,10.5);

\draw[line width=2pt,black,-] (-19.5,7)--(-19.5,10.5)--(-16,10.5);
\draw[fill=white] (-20-0.25,11+0.25) rectangle (-19+0.25,10-0.25);
\draw[fill=white] (-20-0.25,8.5+0.25) rectangle (-19+0.25,7.5-0.25);
\draw[fill=white] (-17.5-0.25,11+0.25) rectangle (-16.5+0.25,10-0.25);

\draw[line width=2pt,black,dotted,-] (-19.5,5.1)--(-19.5,6);
\draw[line width=2pt,black,dotted,-] (-19.5,4.9)--(-19.5,4);

\draw[line width=2pt,black,-] (-19.5,3)--(-19.5,-0.5)--(-16,-0.5);
\draw[fill=white] (-20-0.25,-1-0.25) rectangle (-19+0.25,0+0.25);
\draw[fill=white] (-20-0.25,1.5-0.25) rectangle (-19+0.25,2.5+0.25);
\draw[fill=white] (-17.5-0.25,-1-0.25) rectangle (-16.5+0.25,0+0.25);

\draw[line width=2pt,black,dotted,-] (-13.9,-0.5)--(-13,-0.5);
\draw[line width=2pt,black,dotted,-] (-14.1,-0.5)--(-15,-0.5);

\draw[line width=2pt,black,-] (-12,10.5)--(-8.5,10.5)--(-8.5,7);
\draw[fill=white] (-9-0.25,11+0.25) rectangle (-8+0.25,10-0.25);
\draw[fill=white] (-9-0.25,8.5+0.25) rectangle (-8+0.25,7.5-0.25);
\draw[fill=white] (-11.5-0.25,11+0.25) rectangle (-10.5+0.25,10-0.25);

\draw[line width=2pt,black,dotted,-] (-8.5,5.1)--(-8.5,6);
\draw[line width=2pt,black,dotted,-] (-8.5,4.9)--(-8.5,4);

\draw[line width=2pt,black,-] (-12,-0.5)--(-8.5,-0.5)--(-8.5,3);
\draw[line width=2pt,black,-] (-5,-0.5)--(-8.5,-0.5)--(-8.5,-4);
\draw[fill=white] (-9-0.25,1.5-0.25) rectangle (-8+0.25,2.5+0.25);
\draw[fill=white] (-11.5-0.25,-1-0.25) rectangle (-10.5+0.25,0+0.25);
\draw[fill=white] (-9-0.25,-3.5-0.25) rectangle (-8+0.25,-2.5+0.25);
\draw[fill=white] (-6.5-0.25,-1-0.25) rectangle (-5.5+0.25,0+0.25);

\draw[line width=2pt,black,dotted,-] (-3.1,-0.5)--(-4,-0.5);
\draw[line width=2pt,black,dotted,-] (-2.9,-0.5)--(-2,-0.5);

\draw[line width=2pt,black,-] (-1,-0.5)--(2.5,-0.5)--(2.5,-4);
\draw[fill=white] (-0.5-0.25,0+0.25) rectangle (0.5+0.25,-1-0.25);
\draw[fill=white] (2-0.25,0+0.25) rectangle (3+0.25,-1-0.25);
\draw[fill=white] (2-0.25,-2.5+0.25) rectangle (3+0.25,-3.5-0.25);

\draw[line width=2pt,black,dotted,-] (-8.5,-5.9)--(-8.5,-5);
\draw[line width=2pt,black,dotted,-] (-8.5,-6.1)--(-8.5,-7);

\draw[line width=2pt,black,-] (-8.5,-8)--(-8.5,-11.5)--(-5,-11.5);
\draw[fill=white] (-9-0.25,-8.5+0.25) rectangle (-8+0.25,-9.5-0.25);
\draw[fill=white] (-9-0.25,-11+0.25) rectangle (-8+0.25,-12-0.25);
\draw[fill=white] (-6.5-0.25,-11+0.25) rectangle (-5.5+0.25,-12-0.25);

\draw[line width=2pt,black,dotted,-] (-3.1,-11.5)--(-4,-11.5);
\draw[line width=2pt,black,dotted,-] (-2.9,-11.5)--(-2,-11.5);

\draw[line width=2pt,black,-] (-1,-11.5)--(2.5,-11.5)--(2.5,-8);
\draw[line width=2pt,black,-] (2.5,-12.5)--(2.5,-11.5)--(3.5,-11.5);
\draw[fill=white] (2-0.25,-8.5+0.25) rectangle (3+0.25,-9.5-0.25);
\draw[fill=white] (-0.5-0.25,-11+0.25) rectangle (0.5+0.25,-12-0.25);

\draw[line width=2pt,black,dotted,-] (2.5,-5.9)--(2.5,-5);
\draw[line width=2pt,black,dotted,-] (2.5,-6.1)--(2.5,-7);

\draw[line width=2pt,black,dotted,-] (4.929289321881,-13.929289321881)--(4,-13);
\draw[line width=2pt,black,dotted,-] (5.070710678118,-14.070710678118)--(6,-15);

\draw[line width=2pt,black,-] (7.5,-20)--(7.5,-16.5)--(11,-16.5);
\draw[line width=2pt,black,-] (7.5,-15.5)--(7.5,-16.5)--(6.5,-16.5);
\draw[fill=white] (7-0.25,-18.5+0.25) rectangle (8+0.25,-19.5-0.25);
\draw[fill=white] (9.5-0.25,-16+0.25) rectangle (10.5+0.25,-17-0.25);

\draw[line width=2pt,black,dotted,-] (12.9,-16.5)--(12,-16.5);
\draw[line width=2pt,black,dotted,-] (13.1,-16.5)--(14,-16.5);

\draw[line width=2pt,black,-] (15,-16.5)--(18.5,-16.5)--(18.5,-20);
\draw[fill=white] (15.5-0.25,-16+0.25) rectangle (16.5+0.25,-17-0.25);
\draw[fill=white] (18-0.25,-16+0.25) rectangle (19+0.25,-17-0.25);
\draw[fill=white] (18-0.25,-18.5+0.25) rectangle (19+0.25,-19.5-0.25);

\draw[line width=2pt,black,dotted,-] (18.5,-21.9)--(18.5,-21);
\draw[line width=2pt,black,dotted,-] (18.5,-22.1)--(18.5,-23);

\draw[line width=2pt,black,-] (18.5,-24)--(18.5,-27.5)--(15,-27.5);
\draw[fill=white] (18-0.25,-24.5+0.25) rectangle (19+0.25,-25.5-0.25);
\draw[fill=white] (18-0.25,-27+0.25) rectangle (19+0.25,-28-0.25);
\draw[fill=white] (16.5+0.25,-27+0.25) rectangle (15.5-0.25,-28-0.25);

\draw[line width=2pt,black,dotted,-] (7.5,-21.9)--(7.5,-21);
\draw[line width=2pt,black,dotted,-] (7.5,-22.1)--(7.5,-23);

\draw[line width=2pt,black,-] (7.5,-24)--(7.5,-27.5)--(11,-27.5);
\draw[fill=white] (7-0.25,-24.5+0.25) rectangle (8+0.25,-25.5-0.25);
\draw[fill=white] (7-0.25,-27+0.25) rectangle (8+0.25,-28-0.25);
\draw[fill=white] (9.5-0.25,-27+0.25) rectangle (10.5+0.25,-28-0.25);

\draw[line width=2pt,black,dotted,-] (12.9,-27.5)--(12,-27.5);
\draw[line width=2pt,black,dotted,-] (13.1,-27.5)--(14,-27.5);

\filldraw[black] (-8.5,-0.5) circle (0.5);
\filldraw[black] (2.5,-11.5) circle (0.5);
\filldraw[black] (7.5,-16.5) circle (0.5);
\draw (-19.5,10.5) node {\scriptsize $O$};
\draw (-19.5,8) node {\scriptsize $C$};
\draw (-8.5,-3) node {\scriptsize $C$};
\draw (-8.5,2) node {\scriptsize $F$};
\draw (2.5,-9) node {\scriptsize $F$};
\draw (10,-16.5) node {\scriptsize $C$};
\end{tikzpicture}
\end{center}
\captionshift
\caption{The structure of the highest level of $G_\text{chain}$. Instead of one cycle, we have a series of cycles. In blue the route \opt takes, in red a possible route of \alg.}\label{figure: Basic Ring Structure}
\end{figure}
}
\full{\figureBasicRingStructure}

Let $x,N\in\N$ be large and $x$ be an even number. 
We use the same level-based recursive structure as introduced in \Cref{section: Improved Recursive Construction}.
However, the highest level is not constructed like a cycle, but like a series of connected cycles.
The basic idea is to construct $G_\text{chain}$ in such a way that \alg takes the wrong path at the vertices connecting the cycles.
More precisely, \alg first completely traverses a cycle before moving on to the next one, effectively traversing every cycle twice:
One time by taking the wrong path, half a time to get to the next cycle and another half a time on the way back to the origin.
Every cycle is similarly structured as the macroscopic cycle of \Cref{section: Improved Lower Bound of 3}. 
We describe the graph~$G_\text{chain}$.

\begin{GFour}\label{strategy: Many Circles}
Let $N$ be the highest level.
\alg starts at the origin in an origin block of type~$O^0$, which is (through multiple levels) contained in an origin block of type $O^N$.
Once \alg leaves the origin block, we let \alg enter a normal block of type $B^N$ connected to the origin block via a skip edge as before.
Similarly, every time \alg enters a new block it is a normal block.
The $(\frac{1}{2}x+2)$-nd block \alg examines on one side of the origin block (excluding the origin block) is a final block of type~$F^N$ that is connected to a vertex via its exit edge.
We call the path of blocks between the origin block and the final block ``upper half'' and the path of blocks on the other side of the origin block ``lower half''.
The vertex that is adjacent to the final block is called \emph{connection vertex} $v_\text{con}$ and it is incident to six edges of weight~$e_N$ and one edge of weight $0$ (three skip edges and three backbone edges leading to three unexplored blocks and the exit edge of the final block of weight $0$).
Now \alg has the choice to either take one of the six edges of weight~$e_N$ at $v_\text{con}$ or to backtrack to the origin and explore the blocks on the ``lower half'' further. 

If \alg backtracks to the origin block and explores the lower half, it continues to explore normal blocks until we connect the $(\frac{1}{2}x+1)$-st normal block of type $B^N$ on the lower half to a closing block of type $C^N$.
This closing block is connected to~$v_\text{con}$ by connecting one of the six edges of weight~$e_N$ incident to~$v_\text{con}$ to the tail vertex of the closing block and another one to the pseudo start vertex of the closing block.
See an illustration of this scenario for a cycle that is not the first one on the left side of \Cref{figure: Backtrack Path}.

Now, has examined~$z$ normal blocks on the lower half before it visits~$v_\text{con}$ the first time from the upper half.
If \alg chooses to take one of the six edges of weight~$e_N$ at the connection vertex, we let it enter a normal block of type~$B^N$ and we continue adding normal blocks of type~$B^N$ next to every traversed normal block of type~$B^N$.
Note that~$v_\text{con}$ has six incident edges of weight~$e_N$, which can be used as three pairs of skip and backbone edges.
Therefore we can add up to three paths of normal blocks of type~$B^N$ to~$v_\text{con}$, if \alg chooses to take another edge of weight~$e_N$.
Once \alg has examined $\frac{1}{2}x-z+1$ blocks of type~$B^N$ in one of the three paths, we add a closing block of type~$C^N$.
This block will be connected to the last examined block on the lower half or the origin block (in the case $z=0$).
See an illustration of this scenario with $z=0$ for a cycle that is not the first one on the right side of \Cref{figure: Backtrack Path}.
Note that \alg now needs to backtrack to~$v_\text{con}$ to get to the next cycle.

This procedure can be repeated for every cycle.
Note that apart from the cycle containing the origin, every other cycle is visited by \alg through a connection vertex.
In the last cycle, after \alg has explored a total of $x$ normal blocks on the upper and lower half combined (excluding the connection vertex), we connect both halves with a closing block.
\end{GFour}

As $G_\text{simple}$ and $G_\text{rec}$ before, $G_\text{chain}$ is also a planar graph: 
The argument is the same as before with $G_\text{rec}$, with the slight change that, if we exclude skip and return edges, the resulting graph is a series of cycles instead of just one cycle.

We bound the cost of \alg for traversing one cycle which allows us to prove \Cref{theorem: Lower Bound 4}.
Note that we use the notation introduced in \Cref{section: Improved Recursive Construction} and that $\tX{i}$, $\eX{i}$, $\vX{i}$, $\utX{i}$ and $\uX{i}$ are defined as before.

\newcommand{\figureBacktrackPath}{
\begin{figure}\conf{[p]}
\begin{center}
\begin{tikzpicture}[scale=0.25,font=\scriptsize, rotate=45]

\draw[line width=1pt,red,-,dashed] (-18,12)--(-20,14);
\draw[line width=1pt,red,-] (-18,12)--(-7,12)--(-7,2)--(-10,2)--(-10,9)--(-18,9)--(-18,1)--(-7,1);
\draw[line width=1pt,red,-,dashed] (-7,1)--(-5,-1);
\draw[line width=1pt,red,-,dashed] (-10,-2)--(-8,-4);
\draw[line width=1pt,red,-] (-10,-2)--(-21,-2)--(-21,9);
\draw[line width=1pt,red,->, >=stealth] (-21,9)--(-23,11);

\draw[line width=1pt,red,-,dashed] (-23+20,11-20)--(-21+20,9-20);
\draw[line width=1pt,red,-] (-21+20,9-20)--(-10+20,9-20)--(-10+20,1-20)--(-18+20,1-20)--(-18+20,8-20)--(-21+20,8-20)--(-21+20,-2-20)--(-10+20,-2-20);
\draw[line width=1pt,red,-,dashed] (-7+20,1-20)--(-5+20,-1-20);
\draw[line width=1pt,red,-,dashed] (-10+20,-2-20)--(-8+20,-4-20);
\draw[line width=1pt,red,-] (-7+20,1-20)--(-7+20,12-20)--(-18+20,12-20);
\draw[line width=1pt,red,->, >=stealth] (-18+20,12-20)--(-20+20,14-20);

\draw[line width=2pt,black,dotted,-] (-13.9,10.5)--(-13,10.5);
\draw[line width=2pt,black,dotted,-] (-14.1,10.5)--(-15,10.5);

\draw[line width=2pt,black,dotted,-] (4.929289321881-27,-13.929289321881+27)--(4-27,-13+27);
\draw[line width=2pt,black,dotted,-] (5.070710678118-27,-14.070710678118+27)--(6-27,-15+27);

\draw[line width=2pt,black,-] (-20.5,10.5)--(-16,10.5);
\draw[line width=2pt,black,-] (-19.5,11.5)--(-19.5,10.5);
\draw[line width=2pt,black,-] (-19.5,7)--(-19.5,10.5)--(-16,10.5);
\draw[fill=white] (-20.25,8.75) rectangle (-18.75,7.25);
\draw[fill=white] (-17.75,11.25) rectangle (-16.25,9.75);

\draw[line width=2pt,black,dotted,-] (-19.5,5.1)--(-19.5,6);
\draw[line width=2pt,black,dotted,-] (-19.5,4.9)--(-19.5,4);

\draw[line width=2pt,black,-] (-19.5,3)--(-19.5,-0.5)--(-16,-0.5);

\draw[fill=white] (-20-0.25,-1-0.25) rectangle (-19+0.25,0+0.25);
\draw[fill=white] (-20-0.25,1.5-0.25) rectangle (-19+0.25,2.5+0.25);
\draw[fill=white] (-17.5-0.25,-1-0.25) rectangle (-16.5+0.25,0+0.25);

\draw[line width=2pt,black,dotted,-] (-13.9,-0.5)--(-13,-0.5);
\draw[line width=2pt,black,dotted,-] (-14.1,-0.5)--(-15,-0.5);

\draw[line width=2pt,black,-] (-12,10.5)--(-8.5,10.5)--(-8.5,7);
\draw[fill=white] (-9-0.25,11+0.25) rectangle (-8+0.25,10-0.25);
\draw[fill=white] (-9-0.25,8.5+0.25) rectangle (-8+0.25,7.5-0.25);
\draw[fill=white] (-11.5-0.25,11+0.25) rectangle (-10.5+0.25,10-0.25);

\draw[line width=2pt,black,dotted,-] (-8.5,5.1)--(-8.5,6);
\draw[line width=2pt,black,dotted,-] (-8.5,4.9)--(-8.5,4);

\draw[line width=2pt,black,-] (-12,-0.5)--(-7.5,-0.5);
\draw[line width=2pt,black,-] (-8.5,3)--(-8.5,-1.5);

\draw[fill=white] (-9-0.25,1.5-0.25) rectangle (-8+0.25,2.5+0.25);
\draw[fill=white] (-11.5-0.25,-1-0.25) rectangle (-10.5+0.25,0+0.25);

\draw[line width=2pt,black,dotted,-] (4.929289321881-11,-13.929289321881+11)--(4-11,-13+11);
\draw[line width=2pt,black,dotted,-] (5.070710678118-11,-14.070710678118+11)--(6-11,-15+11);

\draw[line width=2pt,black,dotted,-] (-13.9+20,10.5-20)--(-13+20,10.5-20);
\draw[line width=2pt,black,dotted,-] (-14.1+20,10.5-20)--(-15+20,10.5-20);

\draw[line width=2pt,black,dotted,-] (4.929289321881-7,-13.929289321881+7)--(4-7,-13+7);
\draw[line width=2pt,black,dotted,-] (5.070710678118-7,-14.070710678118+7)--(6-7,-15+7);

\draw[line width=2pt,black,-] (-20.5+20,10.5-20)--(-19.5+20,10.5-20);
\draw[line width=2pt,black,-] (-19.5+20,7-20)--(-19.5+20,11.5-20);
\draw[line width=2pt,black,-] (-19.5+20,7-20)--(-19.5+20,10.5-20)--(-16+20,10.5-20);
\draw[fill=white] (-20+20-0.25,8.5-20+0.25) rectangle (-19+20+0.25,7.5-20-0.25);
\draw[fill=white] (-17.5+20-0.25,11-20+0.25) rectangle (-16.5+20+0.25,10-20-0.25);

\draw[line width=2pt,black,dotted,-] (-19.5+20,5.1-20)--(-19.5+20,6-20);
\draw[line width=2pt,black,dotted,-] (-19.5+20,4.9-20)--(-19.5+20,4-20);

\draw[line width=2pt,black,-] (-19.5+20,3-20)--(-19.5+20,-0.5-20)--(-16+20,-0.5-20);

\draw[fill=white] (-20+20-0.25,-1-20-0.25) rectangle (-19+20+0.25,0-20+0.25);
\draw[fill=white] (-20+20-0.25,1.5-20-0.25) rectangle (-19+20+0.25,2.5-20+0.25);
\draw[fill=white] (-17.5+20-0.25,-1-20-0.25) rectangle (-16.5+20+0.25,0-20+0.25);

\draw[line width=2pt,black,dotted,-] (-13.9+20,-0.5-20)--(-13+20,-0.5-20);
\draw[line width=2pt,black,dotted,-] (-14.1+20,-0.5-20)--(-15+20,-0.5-20);

\draw[line width=2pt,black,-] (-12+20,10.5-20)--(-8.5+20,10.5-20)--(-8.5+20,7-20);
\draw[fill=white] (-9+20-0.25,11-20+0.25) rectangle (-8+20+0.25,10-20-0.25);
\draw[fill=white] (-9+20-0.25,8.5-20+0.25) rectangle (-8+20+0.25,7.5-20-0.25);
\draw[fill=white] (-11.5+20-0.25,11-20+0.25) rectangle (-10.5+20+0.25,10-20-0.25);

\draw[line width=2pt,black,dotted,-] (-8.5+20,5.1-20)--(-8.5+20,6-20);
\draw[line width=2pt,black,dotted,-] (-8.5+20,4.9-20)--(-8.5+20,4-20);

\draw[line width=2pt,black,-] (-12+20,-0.5-20)--(-7.5+20,-0.5-20);
\draw[line width=2pt,black,-] (-8.5+20,3-20)--(-8.5+20,-1.5-20);

\draw[fill=white] (-9+20-0.25,1.5-20-0.25) rectangle (-8+20+0.25,2.5-20+0.25);
\draw[fill=white] (-11.5+20-0.25,-1-20-0.25) rectangle (-10.5+20+0.25,0-20+0.25);

\draw[line width=2pt,black,dotted,-] (4.929289321881-11+20,-13.929289321881+11-20)--(4-11+20,-13+11-20);
\draw[line width=2pt,black,dotted,-] (5.070710678118-11+20,-14.070710678118+11-20)--(6-11+20,-15+11-20);

\filldraw[black] (-8.5,-0.5) circle (0.5);
\filldraw[black] (-19.5,10.5) circle (0.5);
\filldraw[black] (-8.5+20,-0.5-20) circle (0.5);
\filldraw[black] (-19.5+20,10.5-20) circle (0.5);
\draw (-8.5,2) node {\scriptsize $F$};
\draw (-11,-0.5) node {\scriptsize $C$};
\draw (11.5,-18) node {\scriptsize $F$};
\draw (-19.5+20,8-20) node {\scriptsize $C$};
\draw (-8.5-1+0.5,-0.5-1-0.5) node {\tiny $v_\text{con}$};
\draw (-19.5-1-0.5,10.5-1+0.5) node {\tiny $v_\text{con}$};
\draw (-8.5+20-1+0.5,-0.5-20-1-0.5) node {\tiny $v_\text{con}$};
\draw (-19.5+20-1-0.5,10.5-20-1+0.5) node {\tiny $v_\text{con}$};
\end{tikzpicture}
\end{center}
\captionshift
\caption{The path \alg takes if it always decides to backtrack after having traversed a connection block for the first time (left) and the path \alg takes if it does not backtrack, i.e., $z=0$ (right).}\label{figure: Backtrack Path}
\end{figure}
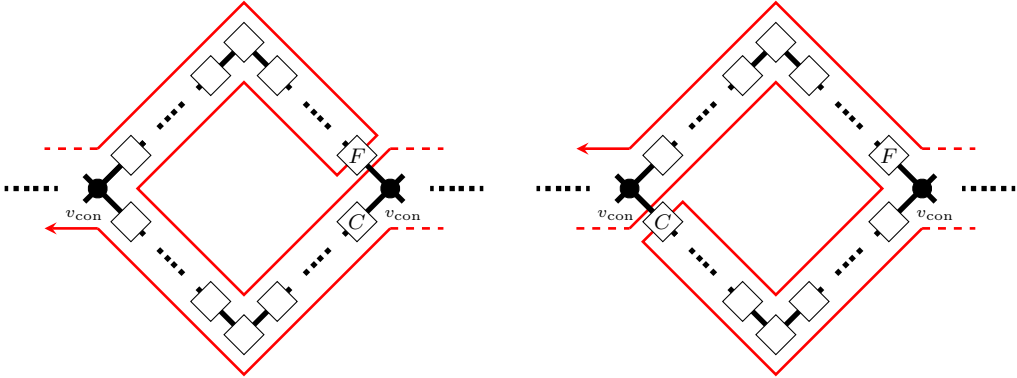
}
\full{\figureBacktrackPath}

\begin{restatable}{Lem}{LemmaCostsPerCircle}\label{lemma: Costs per Circle}
Let $R$ be a cycle in $G_\text{chain}$.
Furthermore, let $R$ not be the last cycle. \alg incurs costs of at least
\begin{equation*}
x(\utX{N}+\tX{N}+\eX{N})
\end{equation*}
inside of $R$ during its exploration of $G_\text{chain}$.
\end{restatable}
\newcommand{\LemmaCostsPerCircleProof}{
\begin{proof}
The cycle $R$ consists of $x+4$ blocks ($x+5$ blocks in case of the first cycle since the first contains an additional origin block). 
If we exclude the closing block of type $C^N$, the final block of type $F^N$ as well as origin blocks of type~$O^N$ in case of the first cycle, exactly~$x+2$ normal blocks remain in any cycle.
Without loss of generality~$R$ is not the first cycle, i.e., it contains two connection vertices instead of one connection vertex and an origin block.
We denote by~$v_1$ the connection vertex that is first reached by \alg and the other by $v_2$.
Note that the proof works similarly for the first cycle by just replacing $v_1$ with an origin block.
It is clear that every normal block of type $B^N$ in $R$ needs to be explored and that \Cref{lemma: Lemma Cost Per Block I} can be applied for at least $\frac{1}{2}x$ of the $\frac{1}{2}x+1$ normal blocks on one half of $R$.
Thus, taking both halves of $R$ into account, \alg accumulates costs of at least $x\utX{N}$ by applying \Cref{lemma: Lemma Cost Per Block I}.
Consider the situation where \alg reaches for~$v_2$ the first time.
We call the path of blocks on the side of~$v_1$ that \alg took to move from~$v_1$ to~$v_2$ upper half. Consequently, the blocks of~$R$ on the other side of~$v_1$ are called lower half.
We distinguish between two cases:
\paragraph*{\textit{Case 1:} \alg chooses to backtrack from~$v_2$ to~$v_1$:}
In this case, \alg accumulates costs of at least $\frac{1}{2}x(\tX{N}+\eX{N})$ by backtracking from~$v_2$ to~$v_1$ and thus traversing at least $\frac{1}{2}x$ already explored normal blocks.
Moving back towards~$v_2$ via the upper half only accumulates additional costs.
Therefore, we may assume that \alg explores the blocks on the lower half and reaches~$v_2$.
At this point \alg has traversed the complete cycle and starts to explore new cycles.
However, at some point it needs to backtrack to the origin to complete its tour.
On its way back, \alg again needs to move from~$v_2$ to~$v_1$ and thus traverse at least $\frac{1}{2}x$ already examined normal blocks again, accumulating a cost of $\frac{1}{2}x(\tX{N}+\eX{N})$.
So, in total, \alg accumulates costs of at least
\begin{equation*}
x(\utX{N}+\tX{N}+\eX{N})
\end{equation*}
as claimed.

\paragraph*{\textit{Case 2:} \alg takes one of the edges of weight $e_N$ at~$v_2$:}
Note that if \alg takes multiple skip edges at $v_2$ it eventually leaves $R$.
Let $z<\frac{1}{2}x$ be the number of normal blocks to the lower half of~$v_1$ that have been explored by \alg.
Since by assumption \alg has reached~$v_2$ via the upper half, \alg has incurred additional costs of $z(\tX{N}+\eX{N})$ on the lower half for backtracking to $v_1$.
Let~$B$ be the last block on the lower half that has been examined by \alg.
By construction, after \alg has left~$v_2$, we add normal blocks of type~$B^N$ until \alg has explored $\frac{1}{2}x-z+1$ of them consecutively after leaving~$v_2$ and then add a closing block of type $C^N$ that is connected to~$B$.
Observe that \alg has accumulated costs of at least
\begin{equation*}
x\utX{N}+z(\tX{N}+\eX{N})
\end{equation*}
until it enters the closing block.
At this point, \alg needs to move back to $v_2$ to find unexplored cycles.
This costs again at least $(\tfrac{1}{2}x-z)(\tX{N}+\eX{N})$ since at least $\tfrac{1}{2}x-z$ already explored normal blocks need to be traversed.
Finally, \alg has to backtrack to the origin at some point, which means it needs to move from $v_2$ to $v_1$ again and thus traverse at least $\frac{1}{2}x$ already examined normal blocks again.
This incurs costs of at least $\tfrac{1}{2}x(\tX{N}+\eX{N})$. 
So, in total, we have
\begin{align*}
x\utX{N}+z(\tX{N}+\eX{N})+(\tfrac{1}{2}x-z)(\tX{N}+\eX{N})+\tfrac{1}{2}x(\tX{N}+\eX{N})=x(\utX{N}+\tX{N}+\eX{N}),
\end{align*}
as claimed.
\end{proof}
}
\full{\LemmaCostsPerCircleProof}

In the following, we set 
\begin{equation}
\ucX{N}:=x\uX{N-1}+3\eX{N}.\label{equation: UCirc}
\end{equation}
Note that we have 
\begin{equation}
\utX{N}\ge\ucX{N}\label{equation: UCirc greater U}
\end{equation}
by \Cref{lemma: Lemma Cost Per Block I}, i.e., $\ucX{N}$ is a lower bound on the costs of \alg for exploring a normal block of type $B^N$.

\begin{restatable}{Lem}{LemmaAnalyzeCosts}\label{lemma: Analyze Costs}
Let $N\ge 0$. We have
\begin{equation*}
\utX{N}+\tX{N}+\eX{N}\ge \ucX{N}+3\eX{N}.
\end{equation*}
\end{restatable}
\newcommand{\LemmaAnalyzeCostsProof}{
\begin{proof}
We have
\begin{align*}
\utX{N}+\tX{N}+\eX{N}&\myoverset{Lem. \ref{lemma: Lemma Cost Per Block I}}{\ge}{38} x\utX{N-1}+(x-y_N)(\tX{N-1}+\eX{N-1})+4\eX{N}+\tX{N}\\
&\myoverset{(\ref{equation: Traversal Cost})}{=}{38} x\uX{N-1}+x(\tX{N-1}+\eX{N-1})+4\eX{N}\\
&\myoverset{(\ref{equation: Edge Cost})}{=}{38} x\uX{N-1}+5\eX{N}\\
&\myoverset{(\ref{equation: UCirc})}{=}{38} \ucX{N}+2\eX{N}.
\end{align*}
\end{proof}
}
\full{\LemmaAnalyzeCostsProof}

The best choice for $y$ in the graph $G_\text{chain}$ is $y=\frac{x}{2}$.
This is the case since \alg is forced to do more backtracking as in $G_\text{rec}$.

\begin{restatable}{Lem}{LemmaRepresentationUTwo}\label{lemma: Representation U Two}
Let $N\ge 0$ and $y=\frac{x}{2}$. We have
\begin{equation*}
\ucX{N}+2\eX{N}= x^{N+1}\uX{-1}+\sum_{j=1}^{N}\tfrac{5}{2}x^j\eX{N-j}+5\eX{N}.
\end{equation*}
\end{restatable}
\newcommand{\LemmaRepresentationUTwoProof}{
\begin{proof}
We show the claim by showing inductively that for every $0\le k \le N$ we have
\begin{equation}
\ucX{N}+2\eX{N} = x^{k+1}\uX{N-k-1}+\sum_{j=1}^k\tfrac{5}{2}x^j\eX{N-j}+5\eX{N}.\label{equation: Induction Three}
\end{equation}
For $k=0$ we have
\begin{equation*}
\ucX{N}+2\eX{N} \overset{(\ref{equation: UCirc})}{=} x\uX{N-1}+5\eX{N}.
\end{equation*}
as claimed.
Now assume the claim is true for $k-1$.
Then we have
\begin{align*}
\ucX{N}+2\eX{N} &\myoverset{(\ref{equation: Induction Three})}{=}{15} x^{k}\uX{N-(k-1)-1}+\sum_{j=1}^{k-1}\tfrac{5}{2}x^j\eX{N-j}+5\eX{N}\\
&\myoverset{(\ref{equation: U General})}{=}{15} x^{k}(x\uX{N-k-1}+\tfrac{5}{2}\eX{N-k})+\sum_{j=1}^{k-1}\tfrac{5}{2}x^j\eX{N-j}+5\eX{N}\\
&\myoverset{}{=}{15} x^{k+1}\uX{N-k-1}+\sum_{j=1}^{k}\tfrac{5}{2}x^j\eX{N-j}+5\eX{N},
\end{align*}
i.e., the equality (\ref{equation: Induction Three}) holds for $k$. 
The claim now follows since (\ref{equation: Induction Three}) also holds for $k=N$.
\end{proof}
}
\full{\LemmaRepresentationUTwoProof}

\begin{restatable}{Lem}{LemmaUTEELargerVE}\label{lemma: UTEE Larger VE}
Let $N\ge 0$ and $y=\frac{x}{2}$. We have
\begin{equation*}
\ucX{N}+2\eX{N}= \left(\frac{10}{3}-\frac{2}{3N+6}\right)\vX{N}-O(x^N),
\end{equation*}
\end{restatable}
\newcommand{\LemmaUTEELargerVEProof}{
\begin{proof}
We have 
\begin{align*}
\ucX{0}+2\eX{0}&\myoverset{(\ref{equation: UCirc})}{=}{15} 6x\\
&\myoverset{}{=}{15} 3(2x+2)-6\\
&\myoverset{}{=}{15} 3\vX{0}-O(1).
\end{align*}
Thus, the claim holds for $N=0$.
Now let $N\ge 1$. We show the claim by showing the equality
\begin{equation}
\ucX{N}+2\eX{N}= x^{N-k}\left(\frac{10}{3}-\frac{2}{3N+6}\right)\vX{k}+\sum_{j=0}^{N-k-1}x^j\left(\frac{10}{3}-\frac{2}{3N+6}\right)\eX{N-j}-O(x^N)\label{equation: Induction Four}
\end{equation}
inductively.
Let $k=0$.
We have
\begin{align*}
\ucX{N}+2\eX{N}&\myoverset{Lem. \ref{lemma: Representation U Two}}{=}{90} x^{N+1}\uX{-1}+\sum_{j=1}^N\frac{5}{2}x^j\eX{N-j}+5\eX{N}\\
&\myoverset{$\uX{-1}=\vX{-1}$}{=}{90} x^{N+1}\vX{-1}+\sum_{j=1}^N\frac{5}{2}x^j\eX{N-j}+5\eX{N}\\
&\mybothset{Lem. \ref{lemma: Two Times Edge}}{=}{$5x\eX{j-1}=\frac{5}{2}x\eX{j-1}+\frac{5}{3}\eX{j}$}{90}{90} x^{N+1}\vX{-1}+5x^N\eX{0}+\sum_{j=0}^{N-1}\frac{10}{3}x^j\eX{N-j}\\
&\mybothset{Lem. \ref{lemma: Costs Easy Opt Final}}{=}{$\vX{0}=x\vX{-1}+\eX{0}+O(1)$}{90}{90} x^N\vX{0}+ 4x^N\eX{0}+\sum_{j=0}^{N-1}\frac{10}{3}x^j\eX{N-j}-O(x^N)\\
&\myoverset{}{=}{90} x^N\vX{0}+x^N\left(4+\frac{2N}{3N+6}\right)\eX{0}\\
&\myoverset{}{}{90}+\sum_{j=0}^{N-1}x^j\left(\frac{10}{3}-\frac{2}{3N+6}\right)\eX{N-j}-O(x^N)\\
&\myoverset{}{=}{90} x^N\vX{0}+x^N\left(\frac{14}{3}-\frac{4}{3N+6}\right)\eX{0}\\
&\myoverset{}{}{90}+\sum_{j=0}^{N-1}x^j\left(\frac{10}{3}-\frac{2}{3N+6}\right)\eX{N-j}-O(x^N)\\
&\myoverset{$\vX{0}=2\eX{0}+O(1)$}{=}{90} x^N\left(\frac{10}{3}-\frac{2}{3N+6}\right)\vX{0}\\
&\myoverset{}{}{90}+\sum_{j=0}^{N-1}x^j\left(\frac{10}{3}-\frac{2}{3N+6}\right)\eX{N-j}-O(x^N)
\end{align*}
as claimed.
Now assume equation (\ref{equation: Induction Four}) is satisfied for $k-1$.
We have 
\begin{align*}
\ucX{N}+2\eX{N}&\myoverset{(\ref{equation: Induction Four})}{=}{95} x^{N-k+1} \left(\frac{10}{3}-\frac{2}{3N+6}\right)\vX{k-1}\\
&\myoverset{}{}{95}+\sum_{j=0}^{N-k}x^j\left(\frac{10}{3}-\frac{2}{3N+6}\right)\eX{N-j}-O(x^N)\\
&\mybothset{Lem. \ref{lemma: Costs Easy Opt Final}}{=}{$\vX{k}=x\vX{k-1}+\eX{k}+O(x^k)$}{95}{95} x^{N-k}\left(\frac{10}{3}-\frac{2}{3N+6}\right)\vX{k}\\
&\myoverset{}{}{95}+\sum_{j=0}^{N-k-1}x^j\left(\frac{10}{3}-\frac{2}{3N+6}\right)\eX{N-j}-O(x^N)\\
\end{align*}
i.e., equation (\ref{equation: Induction Four}) is satisfied for $k$.
The claim now follows from the fact that equation (\ref{equation: Induction Four}) is also satisfied $k=N$.
\end{proof}
}
\full{\LemmaUTEELargerVEProof}

\newcommand{\TheoremLowerBoundFourProof}{
\begin{proof}[Proof of \Cref{theorem: Lower Bound 4}]
For convenience, we only calculate \alg's costs for the first $x$ cycles of $G_\text{chain}$.
Note that this suffices since the true costs of cost of \alg can only be higher.
According to \Cref{lemma: Costs per Circle}, \alg incurs a cost of
\begin{equation*}
x(\utX{N}+\tX{N}+\eX{N})
\end{equation*}
per cycle.
Thus, in total we have
\begin{equation}
\alg(G_\text{chain})\ge x^2(\utX{N}+\tX{N}+\eX{N}).\label{equation: Alg Four}
\end{equation}
$G_\text{chain}$ contains $x+1$ cycles consisting of $x+2$ blocks of level $N$ each (except the first cycle, which contains of $x+3$ blocks).
Thus, $G_\text{chain}$ contains a total of $(x+1)(x+2)+1$ blocks of level $N$.
Thus, \opt explores $(x+1)(x+2)+1$ blocks.
Furthermore, there are $(x-1)x+2(x+2)$ backbone edges of weight $\eX{N}$ to traverse ($x+2$ in the first and the last cycle, $x$ in the remaining $x-1$ cycles).
Thus, there are $2x-1$ more backbone edges to traverse than blocks to explore.
In total, we have
\begin{equation}
\opt(G_\text{chain})=((x+1)(x+2)+1)\vX{N}+(2x-1)\eX{N}=x^2\vX{N}+O(x^{N+2}).\label{equation: Opt Four}
\end{equation}
The competitive ratio is
\begin{align*}
g(x)&\myoverset{}{\coloneqq}{40}\frac{\alg(G_\text{chain})}{\opt(G_\text{chain})}\\
&\myoverset{(\ref{equation: Alg Four}),(\ref{equation: Opt Four})}{\ge}{40}\frac{x^2(\utX{N}+\tX{N}+\eX{N})}{x^2\vX{N}+O(x^{N+2})}\\
&\myoverset{Lem. \ref{lemma: Analyze Costs}}{\ge}{40}\frac{x^2(\uX{N}+2\eX{N})}{x^2\vX{N}+O(x^{N+2})}\\
&\myoverset{Lem. \ref{lemma: UTEE Larger VE}}{=}{40}\frac{x^2\left(\frac{10}{3}-\frac{2}{3N+6}\right)\vX{N}-O(x^{N+2})}{x^2\vX{N}+O(x^{N+2})}\\
\end{align*}
Note that
\begin{equation*}
x^2\left(\frac{10}{3}-\frac{2}{3N+6}\right)\vX{N}\in O(x^{N+3})\quad\text{and}\quad x^2\vX{N}\in O(x^{N+3}).
\end{equation*}
Thus, we get
\begin{align*}
g(x)\ge\frac{10}{3}-\frac{2}{3N+6}\overset{N\rightarrow \infty}{\longrightarrow}\frac{10}{3}>\frac{10}{3}-\varepsilon,
\end{align*}
which completes the proof.
\end{proof}
}
\full{
\TheoremLowerBoundFourProof
}

\section{Conclusion}\label{section: Conclusion}

In this final section we summarize the improvements that have been achieved in the sections above.
Every block constructed in the prior sections is planar, making the complete graph $G$ planar.
Since the competitive ratio for planar graphs is at most 16 \cite{MegowMehlhornSchweitzer/11}, the gap of the competitive ratio of online graph exploration for planar graphs has been narrowed.
\begin{Kor}\label{theorem: General Planar Competitive Gap}
There is no algorithm for online graph exploration with competitive ratio smaller than $\frac{10}{3}$, even for planar graphs.
\end{Kor}
Megow et al.~\cite{MegowMehlhornSchweitzer/11} present an online algorithm for online graph exploration that achieves $2k$-competitiveness in the case that the graph $G$ only has $k$ distinct weights.
By limiting the number~$N$ of recursive levels, our construction yields the following lower bound.
\begin{restatable}{Kor}{TheoremDistinctWeightsPlanarCompetitiveGap}\label{theorem: Distinct Weights Planar Competitive Gap}
There is no algorithm for online graph exploration with competitive ratio smaller than $\frac{10}{3}-\frac{2}{3k}$ for a graph with $k>1$ distinct weights.
\end{restatable}
\newcommand{\TheoremDistinctWeightsPlanarCompetitiveGapProof}{
\begin{proof}
If we have a fixed number $N\ge 0$ of levels, the graph contains $N+3$ distinct weights.
However, it is possible to replace the edges of weight $0$ in $G_\text{chain}$ with edges of weight $1$, reducing the amount of distinct edge weights to $k:=N+2$.
Let this graph be called $G^1_\text{chain}$
This modification increases the costs for \opt for traversing a block of level~$0$ by not more than $4$.
In general, for every block, not more than $0$-weighted edges are replaced with $1$-weighted edges.
Since every block of level $i$ has at most $x+3$ subblocks of level $i-1$ and we have less than $(x+3)^2$ blocks of level $N$, we can bound the number of edges of weight $0$ in graph $G_\text{chain}$ by
\begin{equation*}
4(x+3)^{N+2}\in O(x^{N+2}).
\end{equation*}
We have
\begin{align*}
h(x)&\coloneqq\frac{\alg(G^1_\text{chain})}{\opt(G^1_\text{chain})}\\
&\phantom{:}\ge\frac{\alg(G_\text{chain})}{\opt(G_\text{chain})+O(x^{N+2})}
\end{align*}
Asymptotically, we get
\begin{align*}
h(x)&\overset{x\rightarrow \infty}{\longrightarrow}\frac{10}{3}-\frac{2}{3N+6}=\frac{10}{3}-\frac{2}{3k},
\end{align*}
which shows the claim.
\end{proof}
}
\full{\TheoremDistinctWeightsPlanarCompetitiveGapProof}

\bibliographystyle{abbrv}
{\small\bibliography{references}}

\conf{
\renewcommand{\conf}[1]{}
\cleardoublepage

\FigureSimpleCircleOverview
\figureNormalBlock
\figureOriginBlock
\figureCompatibility
\figureThreeLevels
\figureOriginBlockO
\figureFinalBlockO
\figureChainLevelIEasierNotation
\figureNormalBlockI
\figureFinalBlockI
\figureFinalBlockN
\figureBasicRingStructure
\figureConnectionBlock
\figureBacktrackPath
\cleardoublepage
\appendix
\section*{Appendix} \label{section: Appendix}
\LemmaCostEasy*
\LemmaCostEasyProof

\LemmaCostPerBlock*
\LemmaCostPerBlockProof

\TheoremLowerBoundTwo*
\TheoremLowerBoundTwoProof

\LemmaCostEasyLevelO*
\LemmaCostEasyLevelOProof

\LemmaTraversalCosts*
\LemmaTraversalCostsProof

\LemmaCostEasyLevelI*
\LemmaCostEasyLevelIProof

\LemmaCostsEasyRecursive*
\LemmaCostsEasyRecursiveProof


\LemmaCompatibilityRecursiveN*
\LemmaCompatibilityRecursiveNProof

\LemmaCostPerBlockI*
\LemmaCostPerBlockIProof

\LemmaDegreeOfPols*
\LemmaDegreeOfPolsProof


\LemmaTwoTimesEdge*
\LemmaTwoTimesEdgeProof

\LemmaRepresentationUOne*
\LemmaRepresentationUOneProof

\LemmaUELargerVE*
\LemmaUELargerVEProof

\TheoremLowerBoundThree*
\TheoremLowerBoundThreeProof

\LemmaCostsPerCircle*
\LemmaCostsPerCircleProof

\LemmaAnalyzeCosts*
\LemmaAnalyzeCostsProof

\LemmaRepresentationUTwo*
\LemmaRepresentationUTwoProof

\LemmaUTEELargerVE*
\LemmaUTEELargerVEProof

\TheoremLowerBoundFour*
\TheoremLowerBoundFourProof

\TheoremDistinctWeightsPlanarCompetitiveGap*
\TheoremDistinctWeightsPlanarCompetitiveGapProof
}
\end{document}